\documentclass[preprint]{ametsocjmk}
\usepackage{graphicx}
\usepackage{amsmath}
\usepackage{hyperref}
\usepackage{natbib}

\providecommand\bcdot{\boldsymbol{\cdot}}

\providecommand\boldsymbol[1]{\mbox{\boldmath $#1$}}

\providecommand\bcdot{\boldsymbol{\cdot}}

\newcommand\Cb{\mathbf{C}}

\newcommand\eb{\mathbf{e}}

\newcommand\ub{\mathbf{u}}

\newcommand\kb{\mathbf{k}}

\newcommand\Ub{\boldsymbol{U}}

\newcommand\hu{\widehat{u}}

\received{October 2008} \revised{} \accepted{} \journalid{JPO-XXXX}
\authoraddr{Fabrice Ardhuin, Service Hydrographique et Oc\'{e}anographique de la Marine, 29609 Brest, France
\\E-mail: ardhuin@shom.fr}
\slugcomment{{\bf Fabrice's Draft:} \today}

\begin{document}

\title{Observation and estimation of {L}agrangian, Stokes and {E}ulerian currents \\
induced by wind and waves at the sea surface}

\author[1]{Fabrice Ardhuin}
\author[2]{Louis Mari{\'e}}
\author[2]{Nicolas Rascle}
\author[3]{Philippe Forget}
\author[4]{Aaron Roland}
\affil[1]{Service Hydrographique et Oc\'{e}anographique de la
Marine, Brest, France} \affil[2]{Ifremer, Brest, France}
\affil[3]{Laboratoire de Sondages Electromagn\'{e}tiques de
l'Environnement Terrestre, LSEET, Universit\'{e} du Sud
Toulon-Var, France} \affil[4]{Institut fur Wasserbau und
Wasserwirtschaft, Technishe Universitat Darmstadt, Germany}

\begin{abstract}
The surface current response to winds is analyzed in a two-year
time series of a 12~MHz (HF) Wellen Radar (WERA) off the West
coast of France. Consistent with previous observations, the
measured currents, after filtering tides, are of the order of 1.0
to 1.8\% of the wind speed, in a direction 10 to 40 degrees to the
right of the wind, with systematic trends as a function of wind
speed. This Lagrangian current can be decomposed as the vector sum
of a quasi-Eulerian current $\Ub_E$, representative of the top 1~m of the water column, and part of the wave-induced Stokes
drift $\Ub_{ss}$ at the sea surface. Here $\Ub_{ss}$ is estimated with an accurate numerical wave model, thanks to a novel parameterization of wave dissipation processes. Using both observed and modelled wave spectra, $U_{ss}$ is
found to be very well approximated by a simple function of the
wind speed and significant wave height, generally increasing
quadratically with the wind speed. Focusing on a site located 100~km from the mainland, the wave induced contribution of $U_{ss}$ to the radar measurement has an estimated 
magnitude of 0.6 to 1.3\% of the wind speed, in the wind
direction, a fraction that increases with wind speed. The
difference $\Ub_E$ of Lagrangian and Stokes contributions is found
to be of the order of 0.4 to 0.8\% of the wind speed, and 45 to 70
degrees to the right of the wind. This relatively weak
quasi-Eulerian current with a large deflection angle is
interpreted as evidence of strong near-surface mixing, likely
related to breaking waves and/or Langmuir circulations. Summer stratification tends to increase
the $U_E$ response by up to a factor 2, and further increases the
deflection angle of $\Ub_E$ by 5 to 10 degrees. At locations
closer to coast, $U_{ss}$ is smaller, and $U_E$ is larger with a
smaller deflection angle. These results would be transposable to the world ocean if the relative part of geostrophic currents in $\Ub_E$ were weak, which is expected. This decomposition into Stokes drift and
quasi-Eulerian current is most important for the estimation of
energy fluxes to the Ekman layer.  
\end{abstract}

\section{Introduction}
Surface drift constitutes one of the most \vspace{0.3cm} important
applications of the emerging operational oceanography systems \citep[e.g.][]{Hackett&al.2006}, as it
plays an important role in the fate of oil pollutions and larvae recruitment. A quantitative understanding of the relative
contribution of the wave-induced Stokes drift to the near surface
velocities is also paramount for the proper estimation of air-sea
energy fluxes \citep{Kantha&al.2009}. The quantitative variation of surface drift as a
function of the forcing parameters is still relatively poorly
known. In areas of strong currents due to tides or
quasi-geostrophic dynamics, the surface drift current is highly
correlated to the sub-surface current. Otherwise, winds play a
major role in defining the surface velocities.

Recent theoretical and numerical works
\citep{Ardhuin&al.2004b,Kantha&Clayson2004,Rascle&al.2006,Ardhuin&al.2008} have
sought to reconcile historical measurements of Eulerian and
Lagrangian (i.e. drift) velocities with recent knowledge on
wave-induced mixing \citep{Agrawal&al.1992} and wave-induced drift
\citep{Rascle&al.2008}. These suggest that the surface Stokes
drift $U_{ss}$ induced by waves typically accounts for 2/3 of the
surface wind-induced drift, in the open ocean, and that the surface
wind-related Lagrangian velocity $U_L(z)$ is the sum of the strongly
sheared Stokes drift $U_{S}(z)$ and a relatively uniform quasi-Eulerian current
$\hu(z)$, defined by \citet{Jenkins1987} and generalized by
\citet{Ardhuin&al.2008}. The Stokes drift decays rapidly away from the surface on a scale which is the Stokes depth $D_S$.  For deep-water monochromatic waves of wavelength $L$, we take $D_S=L/4$, by analogy with the usual definition of the (twice larger) depth of wave influence for the orbital motion \citep[e.g.][]{Kinsman1965}. Namely, at that depth, the Stokes drift is reduced to 4\% of its surface value. For random waves, a similar result requires a more complex definition, but the approximate same result can be obtained by using the mean wavelength $L_{03}=g T_{m03}^2$ where $T_{m03}$ is the mean period defined from the third moment of the wave frequency spectrum (see Appendix). Smaller values, like $L/(4 \pi)$  used by e.g. \cite{Polton&al.2005}, are more reprensentative of the depth where the Stokes drift is truly significant. 

For horizontally homogeneous conditions, the depth-integrated quasi-Eulerian mass transport vector $\mathbf{M}^m$ is constrained by the balance
between the Coriolis force and the wind ($\tau_a$) and bottom ($\tau_b$) stresses \citep{Hasselmann1970,Ardhuin&al.2004b,Smith2006b}, 
\begin{equation}
 \frac{\partial \mathbf{M}^m}{\partial t}+  \left(\mathbf{M}^m + \mathbf{M}^w\right) \times {\mathbf e}_z=
\tau_{a} - \tau_{b},\label{balance}
\end{equation}
where $ \mathbf{M}^w$ is the (Stokes) mass 'transport'\footnote{Because in the momentum balance (\ref{balance}) the term $\mathbf{M}^w$ drives a component of mean transport that opposes $\mathbf{M}^w$, there is no  net wave-induced transport, except in non-stationary or non-homogenous conditions \cite{Hasselmann1970,Xu&Bowen1994}.} induced by surface gravity waves, $f$ is twice the vertical component of the Earth rotation vector, usually called the 'Coriolis parameter', and ${\mathbf e}_z$ is the vertical unit vector, pointing up. The surface stress vector $\tau_a$ is typically of the order of $\rho_a C_d U_{10}^2$ with $\rho_a$ the air density and $C_d$ in the range 1--2$\times 10^{-3}$ and $U_{10}$ the wind speed at 10~m height. 
The horizontal homogeneity is obviously never achieved strictly \citep[e.g.][]{Pollard1983}, and this aspect will be further discussed in the context of our measurements. 

The wind-driven current is not expected to be significant at a depth greater than 0.7 times the Ekman depth $D_E=0.4 \sqrt(\tau_a/\rho_w)/f$   \citep[i.e. less than 0.2\% of the wind speed if the surface value is 2.8\% of $U_{10}$,][]{Madsen1977}. For a wind speed $U_{10}=10$~m~s$^{-1}$, $0.7 D_E$ is of the order of 30~m. In locations with a larger water depth, the bottom stress is thus expected to be negligible. 
Further, this depth of maximum influence can also be limited by a vertical stratification, with larger velocities in shallow mixed layers, and directions of $U_E$ more strongly deflected to the
right of the wind (in the Northern Hemisphere) than previously
expected \citep{Price&Sundermeyer1999,Rascle2007}. It has also been proposed by \cite{Polton&al.2005} that the wave-induced mass 'transport' $M^w$ may play a role in the modification of near-surface currents, but $M^w$ is generally less than 30\% of the Ekman transport $M^E=\tau_a/f$, and its effect appears to be secondary compared to the stratification \citep{Rascle&Ardhuin2009}. 
The time-averaged balance given by (1) is thus approximately, 
$\mathbf{M}^m=-\mathbf{M}^w+\left(\tau_a \times {\mathbf e}_z\right)/f$. This was nearly verified for the LOTUS3
dataset \citep{Price&Sundermeyer1999}, when allowing for wave-induced biases in the mooring measurements
\citep{Rascle&Ardhuin2009}. Yet, this is not always the case \citep[e.g.][]{Nerheim&Stigebrandt2006}, possibly due to
baroclinic currents and other phenomena that are difficult to separate from the wind-driven component.

The vertical profile of the quasi-Eulerian current is, under the same homogeneous and stationary circumstances, the solution of \citep{Xu&Bowen1994,Ardhuin&al.2008} 
\begin{equation}
 \frac{\partial \widehat{\mathbf{u}}}{\partial t}+  \left(\widehat{\mathbf{u}} + \mathbf{u}_S\right) \times {\mathbf e}_z=
\frac{\partial}{\partial z}\left(K\frac{\partial \widehat{\mathbf{u}}}{\partial z}\right),\label{balance1D}
\end{equation}
where $K$ is a turbulent mixing coefficient. 

These predictions were verified by \cite{Rascle2007} with mooring
data at depths greater than 5~m and surface-following measurements by \cite{Santala&Terray1992} at depths larger than 2~m. When extrapolated to the surface
using a simple numerical model, these observations give directions of $U_E$
between 45$^\circ$ and 90$^\circ$, more than the 45$^\circ$ given
by the constant eddy viscosity model of \citet{Ekman1905}, as
extended by \citet{Gonella1971}, and the 10$^\circ$ given by the
linear eddy viscosity model of \citet{Madsen1977}. This surface
angle, and the magnitude of $U_E$ is also critical for the estimation of the flux of wind
energy to the Ekman layer \citep[e.g.][]{Wang&Huang2004}, or the
analysis of near-surface drifter data 
\citep[e.g.][]{Rio&Hernandez2003,Elipot&Lumpkin2008}. For a better understanding of these questions, it is thus necessary to use 
ocean velocities measured much closer to the surface.

High Frequency (HF) radars can provide such measurements, at depths that depend on their operating frequency. Using a
30 MHz radar, \citet{Mao&Heron2008} made observations that are
also consistent with the idea that the drift current, found to be
2.1\% of the wind speed on average, is the sum of $U_E$ which,
according to their theory, depends quadratically on the wind
speed, and $U_{ss}$ which they estimate to depend linearly on the
wind speed, with a variation according to the fetch.
Unfortunately, their analysis relied on empirical wave estimates
that give large relative errors \citep[of 
the order of 100\%, see
e.g. ][]{Kahma&Calkoen1992,Ardhuin&al.2007}, and a limited range
of wind speeds. Other HF-radar observations give a surface current
of the order of 1.5 to 2.5\% of $U_{10}$ \citep{Essen1993} with 25
to 30 MHz radars. Dobson et al. (1989\nocite{Dobson&al.1989})
also report a ratio of 2.0\% using a 22 MHz radar, and
\cite{Shay&al.2007} report a ratio of 2 to 3\% using a 16 MHz
radar in water depths of 20 to 50 m. These analyses are 
difficult 
to interpret due to the filters applied on time series
to remove motions (tides, geostrophic currents ...) that are not
related to the wind,  and also because of the importance of
inertial oscillations that make the wind- and wave-driven current
a function of the full wind history, and not just a function of
the wind vector at the same time and location.

 In the present paper we extend the previous analyses of HF radar data by independently
 estimating the Stokes drift, using an accurate wave model. We find
that at our deep water\footnote{This means deeper than both the
Stokes depth $D_S$ and the expected Ekman depth $D_E$.} North-East Atlantic site the
quasi-Eulerian current $U_E$ is of the order of 0.6\% of the wind
speed with a direction that is, on average, 60$^\circ$ to the
right of the wind. We also find that the time-dependent response of surface current to the wind is typical of a slab layer with a transfer function proportional to $1/{(f+\omega)}$, where $\omega$ is the radian frequency considered. This result is expected to be representative of
the open ocean. Therefore the estimates of the flux of wind energy
to the Ekman layer by e. g. \citet{Wang&Huang2004} may not be quantitatively correct: they used an angle of
45$^\circ$, a surface velocity which is $2\sqrt{\tau_a/\rho_w}$ for steady winds (about 0.2\% of the wind speed), and a transfer function proportional to $1/\sqrt{f+\omega}$. A proper
analysis of the effects of waves is needed to properly evaluate energy fluxes. 

Our new data and its processing
are described in section 2, and the analysis of the stratification
effect is presented in section 3 with conclusions in section 4.

\section{Lagrangian and quasi-Eulerian current from HF radars}
\subsection{Radar measurements and processing}
High frequency radars measure, among other things
\citep[e.g.][]{Ivonin&al.2004}, the phase velocity $C$ of Bragg
waves that have a wavelength equal to one half of the radar
electromagnetic wavelength and that propagate in directions away
from and toward the radar. This phase velocity is a combination of
the quasi-Eulerian current $U_{E}$
\citep{Stewart&Joy1974,Kirby&Chen1989},
 the phase speed of linear waves $C_{\mathrm{lin}}$, and a
nonlinear wave correction \citep{Weber&Barrick1977} that can be
interpreted as a filtered surface Stokes drift $U_{Sf}$. For
monostatic systems, the usual radial current velocity in the
direction $\theta_B$ towards one radar can be expressed as
\begin{eqnarray}
U_R(\theta_B) &=& C(\theta_B)-\Cb_{\mathrm{lin}} \bcdot
\eb_{\theta_B}  \nonumber \\
&=& U_{Sf}(\theta_B) + \Ub_{E}\bcdot
\mathbf{e}_{\theta_B},\label{Ur}
\end{eqnarray}
where $\mathbf{e}_{\theta_B}$ is the unit vector in direction
$\theta_B$. This velocity can be loosely interpreted as the
projection in direction $\theta_B$ of a current vector $\Ub_R$.
The reason why this is not exactly true is that $U_{Sf}(\theta_B)$
for all directions cannot be exactly given by the projection of a
vector $\Ub_{Sf}$. In other words, $U_{Sf}(\theta_B)$ is not
exactly proportional to $\cos(\theta_B)$, although it is a
reasonable approximation \citep{Broche&al.1983}.

In order to express $U_{Sf}$, we first define the Stokes drift
vector for waves with frequencies up to $f_c$ from the directional
wave spectrum $E(f,\theta)$,
\begin{equation}
\Ub_{ss}(f_c) = 4 \pi \int_0^{f_c} \int_0^{2 \pi}   f
\kb(f,\theta) E(f,\theta) \mathrm{d} f\mathrm{d} \theta,  \label{Uss}
\end{equation}
where $k(f)$ is the magnitude of the wavenumber $\kb$, equal to $(2\pi f)^2/g$ for
linear waves in deep water,  and $g$ is the acceleration of
gravity. Starting from the full expression given by
\cite{Weber&Barrick1977}, \cite{Broche&al.1983} showed that the
filtered Stokes drift component that affects the radial current
measured by one radar station is well approximated by
\begin{eqnarray}
U_{Sf}(k_B,\theta_B) &\simeq&  \Ub_{ss}(f_B) \bcdot
\mathbf{e}_{\theta_B}   \nonumber \\
& +& 4 \pi  k_B \int_{f_B}^\infty \int_0^{2 \pi} f
\cos(\theta-\theta_B) E(f,\theta) \mathrm{d} \theta \mathrm{d}
f\label{Usf}  \nonumber \\
\end{eqnarray}
where $f_B$ is the frequency of the Bragg waves, and $\kb_B$ is
the corresponding wavenumber vector, with a direction $\theta_B$
and magnitude $k_B$. The full expression, correcting typographic
errors in \cite{Broche&al.1983} is given in Appendix A. In order
to simplify the notations, the variable $k_b$ in $U_{Sf}$ will now
be omitted, but the filtered Stokes drift is always a function of
the Bragg wavenumber, thus being different for different radar
frequencies.

The depth-varying quasi-Eulerian current $\hu(z)$ is
defined as the difference of the Lagrangian velocity and Stokes
drift \citep{Jenkins1987}, and can generally be estimated from the
full velocity field using a Generalized Lagrangian Mean
\citep{Ardhuin&al.2008}. The value $\Ub_{E}$ estimated from the
radar is, according to linear wave theory, the integral of
$\hu(z)$ weighted by the Bragg wave Stokes drift profile
\citep{Stewart&Joy1974,Kirby&Chen1989}. In deep water this is,
\begin{equation}
\Ub_{E} = 2 k_B  \eb_{\theta_B} \bcdot  \int_{-\infty}^{0} 
\widehat{\ub} \mathrm{e}^{2 k_B z} \mathrm{d} z.  \label{UE}
\end{equation}

 Here we use data from a WERA HF-radar system
\citep{Gurgel&al.1999}, manufactured by Helzel GmbH, and operated at 12.4 MHz. The
Bragg wavelength is 12.1~m, corresponding to a wave frequency of
0.36~Hz in deep water. Thus half of the weight $\mathrm{e}^{2 k_B z}$ in eq. (\ref{UE}) comes from water depths less than 0.6~m from the moving sea surface,
compared to 0.28~m with the 30~MHz radar of \citet{Mao&Heron2008}.
The relative contributions from deeper layers to $U_E$ decrease
exponentially with depth as $\exp(2 k_B z)$. Therefore $U_E$ can be interpreted as the quasi-Eulerian 
current in the top 1~m of the ocean. 

The radar system has been deployed and operated by Actimar SAS,
since July 2006 on the west coast of France  (figure 1), measuring
surface currents and sea states every 20 minutes. The area is
characterized by intense tidal currents, in particular between the
largest islands where it exceeds 3~m~s$^{-1}$ during mean spring
tides. Also important, the offshore stratification is largely
suppressed by mixing due to the currents in the areas shallower
than 90~m, resulting in complex temperature fronts that are
related to the bottom topography
\citep[e.g.][]{Mariette&LeCann1985}.

Each radar station transmits a chirped continuous wave with a
repetition frequency of 4 Hz and a 100~kHz bandwidth which gives a
radial resolution of 1.5~km.  The receiving antennas are 16-element
linear arrays with a spacing of 10 m, giving a typical angular
resolution of 15 degrees. The raw data is processed to remove
most of the interference signals \citep{Gurgel&Barbin2008}.
Ensemble-averaging over 4 consecutive segments of 512 pulses
yields a velocity resolution $d_u=0.09$ m/s in the Doppler spectrum used to estimate each individual
radial current measurement. Yet, the current value is obtained by a weighted sum over a 9-point window applied to the Doppler spectrum. Provided that some inhomogeneity exists in the current field, the width of the Doppler spectrum permits a measurement resolution that is infinitely small, but with an accuracy that is difficult to define, because no other instrument, except maybe for the CODE-type drifter \citep{Davis1985}, 
is able to measure surface current in the top one meter of the ocean. Similarly, satellite altimeters are reported to measure the mean sea level position with an accuracy of the order of 2~cm whereas their typical range resolution is close to 40~cm. \cite{Prandle1987} used the coherence of the tidal motions to infer that the accuracy of his 27~MHz radar system was indeed less than the Doppler resolution when averaged over one hour. We will thus take the accuracy to be equal to the resolution, but as it will appear below, the only source of concern for our analysis is not so much the random error but a systematic bias, since we will average a very large number of independent measurements. 

\begin{figure}[htb!]
\centerline{\noindent\includegraphics[width=\linewidth]{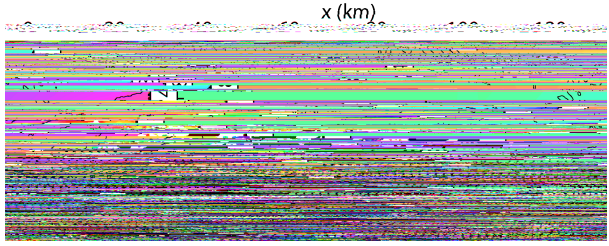}}
%\centerline{\noindent\includegraphics[width=0.7\linewidth]{Iroisemap.pdf}}
\caption{Map of the area showing a map of significant wave height
on January 1st 2008, at 12:00 UTC, estimated with a numerical wave
model (see Appendix B), and the instantaneous surface current
measured by the H.F. radars installed at Porspoder and
Cl{\'e}den-Cap-Sizun. In situ measurement stations include the
weather buoy {BEA}trice and the Pierre Noires (62069) directional
Datawell waverider buoy (installed from November 2005 to March
2006 and back again since January 2008), and a previous waverider
deployment (Iroise), more representative of the offshore wave
conditions. The large black square around point A is the area over
which the radar data has been compiled to provide the time series
analyzed here, representative of offshore conditions. When the radar functionned, over the entire square measurements are available for more than 80\% of the 20 minute records, a number than rises to 99\% for the area East of 5$^\circ$35'W. The partial radar coverage around point A is typical of high sea states with $H_s> 6$~m offshore, which are rare events.\label{map}} 
\end{figure}

Because we investigate the relationship between surface currents and winds based on modelled winds and waves, we will consider only the temporal evolution of the wave field at
one point of the radars' field of view that is representative of
the offshore conditions, at a distance of 80 to 100~km from shore and with a water depth of 120~m. The reason for chosing this location is that we have verifed the wind and wave model results to be most accurate offshore where they were verified in situ with measurements that only span 6 and 9 months of our radar time series. Other reasons for looking at offshore conditions are the expected limited effect of the bottom, and the expected small horizontal gradients of both tidal currents and other processes. Namely, we stay away from the thermal front that typically follows the 90~m depth contour \citep{Mariette&LeCann1985,LeBoyer&al.2009}. The down side of this choice is that the HF-derived current is generally less accurate as the distance from the coast increases, and the coverage is not permanent, especially during severe storms (e.g. figure \ref{map}). These two drawbacks are limited in practice, as we now discuss.

Interferences and ships cause some data to be rejected in the
radar processing, or yield bad measurements, and heavy seas or
calm seas also reduce the working radar range. In order to obtain
a nearly continuous time series, we compiled and filtered data
from a 0.2$^\circ$ in latitude by 0.3$^\circ$ in longitude box
around that point (A in figure 1, the arrow spacing indicate the
resolution of the radar grid). This compilation was done in two
steps. First, based on a visual inspection of the data, at each
radar grid point, 0.05\% of the total number of data points in the
radial velocities time-series are considered spurious and removed.
These points are selected as the points where the raw radial
current time-series differs most from the result of a 5-point
median filter. The 0.05\% value was selected as a convenient
rule-of-thumb, which removes most of the visibly
spurious points, but does not introduce too many unnecessary gaps
in the time-series. Second, the time-series of all the grid points
in the box around A were converted to $u$ and $v$ components and
averaged.

 The Cartesian components of $\Ub_R$ and $\Ub_E$ with
respect to west-east (component $u$) and south-north ($v$)
directions are calculated from the two radial components $U_R(\theta_{B1})$ and $U_R(\theta_{B2})$, 
each measured by one radar station, before and after the substraction of
$\Ub_{Sf}(\theta_B)$. These Cartesian components suffer from a
geometrical dilution of precision (GDOP), varying with position
\citep{Chapman&al.1997,Shay&al.2007}. The radar beams intersect at
point A with an angle  $r= 34^\circ$ and it is possible to
estimate the GDOP values for $u$ and $v$, i.e. the ratios $S_u /s$
and $S_v /s$ where $S_u$, $S_v$ and $S$ are the uncertainties in $u$, $v$
and $u_r$, respectively. Assuming that $S$ has no bias and is
uniformly distributed from $-d_u/2$ to $+d_u/2$, each radar
measurement has an intrinsic uncertainty $S_u=0.04$~m~s$^{-1}$ and
$S_v=0.11$~m~s$^{-1}$.

 This compiled time series, extending
from July 5 2006 to July 31 2008, is the basis of the following
analysis. The 1200~s resolution data was averaged over 3~h blocks
centered on round hours. Gaps shorter than 6~h were linearly
interpolated. That time series is 97\% complete, and thus covers
two full years. Other parts of the radar field of view yield
similar results, briefly discussed below. Due to the averaging in
space and time, each point in the time series is the combination
of about 30 range cells and 9 time intervals, i.e. 180 independent
velocity measurements when the full radar range is obtained. Even with a 11~cm~s$^{-1}$ uncertainty on the original measurement, the expected
r.m.s. error on the velocity components are thus less than
1~cm~s$^{-1}$. This analysis assumes that the instrument is not biased. After verification of the radar antenna lobe patterns using both in situ transmitters and a novel technique based on the analysis of radio interference (to be described elsewhere), the main lobe of the radar is known to be mispointed by less than 5 degrees, with a -3dB width less than 15$^\circ$. The largest source of uncertainty is thus the interpretation of the phase speed and the numerical estimation of the Stokes drift, as discussed below.

Because we wish to focus on the random wind-driven currents, we
also performed a tidal analysis using the T-TIDE software
\citep{Pawlowicz&al.2002} applied to each velocity component. This analysis
on the full time series (before time averaging) allows the removal
of the deterministic diurnal constituents $K_1$, $O_1$, $P_1$ and
$Q_1$ that have amplitudes of 1.5 to 0.3~cm~s$^{-1}$, with
estimated errors of 0.1~cm~s$^{-1}$. Because this only corrects
for 95\% of the apparent variance in the $M_2$ and $S_2$
semi-diurnal tides, these will be further filtered using a time
filter.

%%%%%%%%%%%%%%%%%%%%%%%%%%%%%%%%%%%%%%%%%%%%%%%%%%%%%%%%%%%%%%%%%%%%%%%%%%%%
%%%%%%%%%%%%%%%%%%%%%%%%%%%%%%%%%%%%%%%%%%%%%%%%%%%%%%%%%%%%%%%%%%%%%%%%%%%%%%%

\subsection{Numerical wave model and estimations of Stokes drift}
\subsubsection{General principles}
As expressed by eq. (\ref{Usf}), the estimation of
$U_{Sf}(\theta_B)$ requires the measurement or modelling of the
wave spectrum $E(f,\theta)$. In situ buoys were moored for
restricted periods at several locations for the investigation of
offshore to coastal wave transformation \citep{Ardhuin2006a} and
to provide complementary data for radar validation. The radar also
measures the sea state, but the coverage is often limited, and its
accuracy for a 20 minute record is typically only of the order of
25\% for the significant wave height $H_s$. Thus, in order to use
the full current time series at the offshore location (point A) we
have to estimate the sea state using a numerical wave model.

We use an implementation of the WAVEWATCH III code, in
its version 3.14 \citep{Tolman2007,Tolman2008}, with minor
modifications of the parameterizations, see appendix B, and the
addition of advection schemes on unstructured grids \citep{Roland2008}.

 The model setting consists of a two-way nested pair of
grids, covering the global ocean at 0.5 degree resolution and the
Bay of Biscay and English channel at a resolution of 0.1 degree. A
further zoom over the measurement area is done using an
unstructured grid with 8429 wet points (figure 1). The model
setting is fully described in appendix B.

In practice, $U_{Sf}$ is dominated by the first term
$U_{\mathrm{ss}}(f_B)$, in eq. (\ref{Usf}). Examining a large
number of spectral data (6 buoys for 2 years spanning a range of
wave climates, see appendix C), we realized that
$U_{\mathrm{ss}}(f_B)$ is essentially a function of the wind speed
$U_{10}$ and the wave height $H_s$. While $U_{10}$ explains
typically only 50\% of the variance of $U_{\mathrm{ss}}(f)$ with
$0.3 <  f < 0.5$, $U_{10}$ and $H_s$ generally explain over 85\%
of the variance. This behaviour of $U_{\mathrm{ss}}(f)$  is
similar to that of the fourth spectral moment, related to the
surface mean square slope
\citep{Gourrion&al.2002,Vandemark&al.2004}. The reason for this
correlation is that the wind speed is obviously related to the
high frequency part of the wave spectrum, which determines most of
the Stokes drift, while $H_s$ is a surrogate variable for both the
presence of swell and the stage of development of the wind sea.
Here we find,
\begin{eqnarray}
U_{\mathrm{ss}}(f_c)&\simeq& 5.0\times 10^{-4}
\left[1.25-0.25\left(\frac{0.5}{f_c}\right)^{1.3}\right] U_{10}
\nonumber \\
&  \times& \min\left\{U_{10},14.5\right\} + 0.025\left(
H_s-0.4\right)\label{Uss_U10}.\nonumber\\
\end{eqnarray}
The relationship given by eq. (\ref{Uss_U10}) appears to be very
robust, with a 2.6~cm~$^{-1}$ r. m. s. difference compared to
global hindcast values of $U_{ss}(\infty)$, which is a 16.9\%
difference. Nevertheless, when compared to buoy data, an accurate
wave model generally provides a better fit to the observations
(Appendix C). We thus have used our hindcasts using WAVEWATCH III to provide an estimate for $U_{Sf}$.

\subsubsection{Uncertainty on $U_{Sf}$ around point A}
We have no wave measurement at point A, and no permanent spectral
measurement in the area. A detailed validation of $U_{ss}$ was
thus performed for the coastal buoys 62069 (figure 1), 62064 (off Cap Ferret, 600 km to the southeast of point A), the U.S. Northwest Pacific Coast (appendix C), U.S. East coast, Gulf of Mexico and California.

We further use wave information at buoy
62163, located 150 km west of point A, reprensentative of the offshore conditions found at point A, and a combination of satellite altimeter data. 
The present model estimates of $H_s$ are more accurate at buoy
62163, located 150 km west of point A, than at Pacific buoy
locations. Further, the model estimate of the fourth moment $m_4$
of the wave spectrum is better correlated in the Bay of Biscay to
radar altimeter C-band cross-section, compared to other regions of
the world ocean (Appendix C). We thus expect the model estimate of
$U_{ss}(f_B=0.36~$Hz) to have a bias smaller than than 5\%, with a
random error less than 20\% (see Appendix C). As a result, We
chose to use this numerical wave model for the estimation of
$U_{ss}$ and $U_{Sf}$. 
We can thus propose an error buget for our estimate of the wind-driven quasi-Eulerian current in which the measurement error is dominated by $U_{Sf}$ with a bias of 5\% at most and a standard deviation less than 20\% overall. Using the analysis of 2 years of model results, this standard deviation at the Pacific buoy 46005 is 24\% for wind speeds of 3~m~s$^{-1}$, 20\% for 5~m~s$^{-1}$, 16\% for 7~m~s$^{-1}$, 11\% for 11~m~s$^{-1}$. Given the general accuracy of the wave model in the North-East Atlantic, we expect similar results here. 

We thus estimate that the root mean square error of the modelled quasi-Eulerian current $U_E$ at 3~hour intervals is of the order of 0.2\% of $U_{10}$. On this time scale, it is difficult to rule out contributions from horizontal pressure gradients in the momentum balance, and this current may not be purely wind-driven. 

The averaged current, e.g. for a given class of wind speed, as shown on figure 7, has a relative accuracy better than 0.1\% of $U_{10}$.  In-situ measurements of time-averaged velocities from 10 to 70~m above the bottom at 48$^\circ$6'N and 5$^\circ$23'W  (south of point A, see figure 1) using a RDI Workhorse ADCP deployed from June to September 2007 \citep{LeBoyer&al.2009} give tide-filtered currents less than 2~cm~s$^{-1}$ or 0.25\% of the wind speed when averaged following the wind direction (the instantaneous measurements are rotated before averaging), and less than 0.1\% when winds stronger than 10~m~s$^{-1}$. This is typically less than 20\% of $U_{Sf}$.  Assuming that wind-correlated baroclinic currents are negligible during the ADCP measurement campaign, the wind-correlated geostrophic current is expected to be less than 0.2\% of $U_{10}$. Gereralizing this result to the entire radar time series, the averaged values of $U_E$ can be interpreted as a wind-driven current with an accuracy to within 0.3\% of $U_{10}$.

\section{Analysis of wind-driven flows}
\begin{figure}[htb!]
\centerline{\noindent\includegraphics[width=\linewidth]{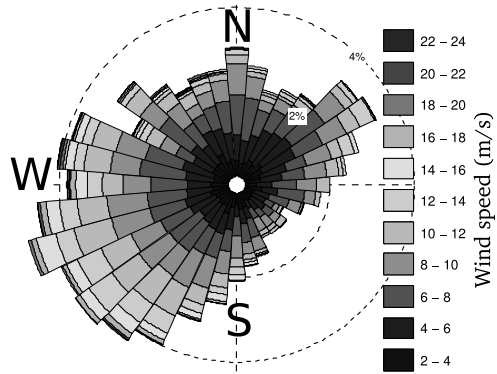}}
%\centerline{\noindent\includegraphics[width=0.7\linewidth]{wind_rose_bw.pdf}}  
\caption{Wind rose for the years 2006 to 2008 at point A, based on ECMWF analyses. The observations at BEAtrice buoy give a similar result. For each direction, the cumulative frequency is indicated with wind speeds increasing from the center to the outside, with a maximum of 4.3\% maximum from West-South-West (heading 250$^\circ$). An isotropic distribution would have a maximum of 2.7\%. \label{windrose}} 
\end{figure}

The study area is characterized dominated by moderate 6 to 12~m~s$^{-1}$  winds, from a wide range of directions, with slightly dominant South-Westerly and North-Easterly sectors (figure \ref{windrose}). 
%%%%%%%%%%%%%%%%%%%%%%%%%%%%%%%%%%%%%%%%%%%%%%%%%%%%%%%%%%%%%%%%%%%%%%%%%%%%
%%%%%%%%%%%%%%%%%%%%%%%%%%%%%%%%%%%%%%%%%%%%%%%%%%%%%%%%%%%%%%%%%%%%%%%%%%%%%%%

\subsection{Rotary spectral analysis}
The rotary spectral analysis gives both the frequency distribution
of the signal, and an indication of its circular polarization
\citep{Gonella1971}. The positive frequencies correspond to
counter-clockwise motions, and the negative frequencies correspond
to clockwise motions, the usual polarization of inertial motions
in the Northern Hemisphere.

 The instantaneous measurements of the radar are dominated by
tidal currents, and the variance of motions with frequencies less
than 1.75 count per day (cpd) only accounts for 8\% of the total
variance (figure 3). These low frequency motions include the
diurnal tidal constituents, most importantly $K_1$ and $O_1$, but
these only account for 0.1\% of the variance. The low
frequency motions are generally dominated by near-inertial
motions, which are polarized clockwise with frequencies close to
the inertial frequency $f_I=1.3$ counts per day (c.p.d., see figure 3).

\subsection{Co-spectral analysis}
Here we investigate the relationship between measured currents,
processed as described above, and winds, taken from 6-hourly wind
analyses from ECMWF. These analyses were verified to give
excellent correlation ($r \simeq 0.92$) with the BEA buoy (WMO
code 62052), which unfortunately malfunctionned during large
periods of time. The wind and current data are thus completely
independent. The wave model was forced by these same winds, and
thus the high level of coherence between the predicted Stokes
drift and the wind (figure 4) is not surprising.

 In order to isolate the wind-correlated dynamics from the shorter (tide) and longer (general
circulation) time scales, we first perform a co-spectral analysis
of the measured currents with the wind, following the method of
\cite{Gonella1971}. In order to keep as much data as possible
between data gaps, the Fourier transforms are taken over 264
hours, which corresponds to 21 $M_2$ tidal cycles. The measured
currents are significantly coherent with the wind vector over the
range -1.75 to 1.75 cpd (figure 4). This coherence is generally
reduced when the Stokes component $U_{Sf}$ is subtracted from the
radar measurements.
%%%%%%%%%%%%%%%%%%%%%%%%%%%%%%%%%%%%%%%%%%%%%%%%%%%%%%%%%%%%%%%%%%%%%%%%%%%%
\begin{figure}[htb!]
\centerline{\noindent\includegraphics[width=\linewidth]{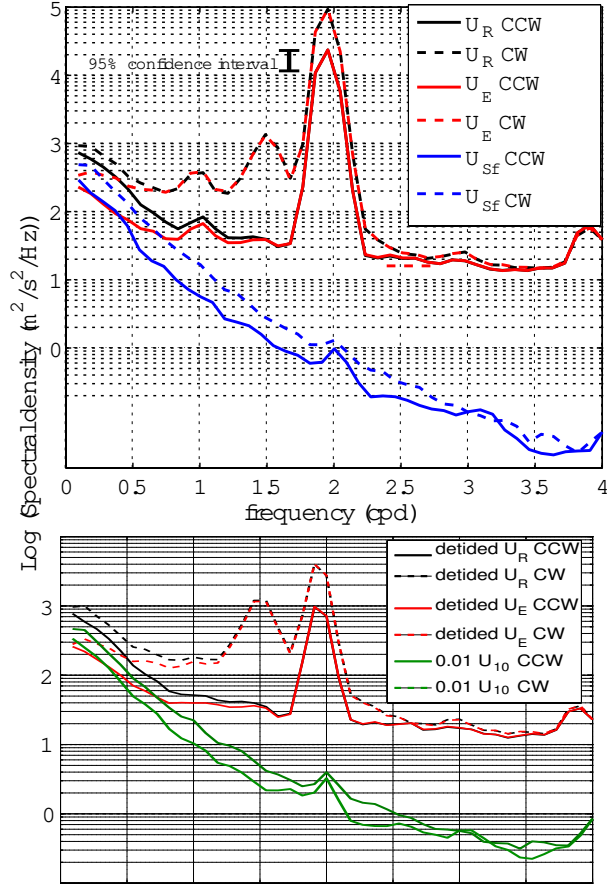}}
%\centerline{\noindent\includegraphics[width=0.7\linewidth]{spectrum_86_revised.pdf}}
\caption{Rotary power spectra of the current measured by the
radar, and the contribution $\Ub_{Sf}$ to the surface Stokes drift
estimated via eq. (\ref{UsfA2}). Clockwise (CW) motions are shown
with dashed lines and counter-clockwise motions are shown with
solid lines. The spectra were estimated using half-overlapping
segments 264~h long over the parts of the time series with no
gaps. The number of degrees of freedom is taken to be the number
of non-overlapping segments, i.e. 59, at the spectral resolution
of 0.09 cpd, giving a relative error of 35\% at the 95\%
confidence level. In the bottom panel the the tidal components have been filtered out, which clearly removes the diurnal peak However, the the semi-diurnal tides are only reduced by a factor 25, 
which is not enough compared to the magnitude of the near-intertial motions, and requires the use of 
an additional filter. This
tide-filtered time series is used in all of the following.}
\label{spec}
\end{figure}
%%%%%%%%%%%%%%%%%%%%%%%%%%%%%%%%%%%%%%%%%%%%%%%%%%%%%%%%%%%%%%%%%%%%%%%%%%%%%%%
%%%%%%%%%%%%%%%%%%%%%%%%%%%%%%%%%%%%%%%%%%%%%%%%%%%%%%%%%%%%%%%%%%%%%%%%%%%%
\begin{figure}[htb!]
\centerline{\noindent\includegraphics[width=\linewidth]{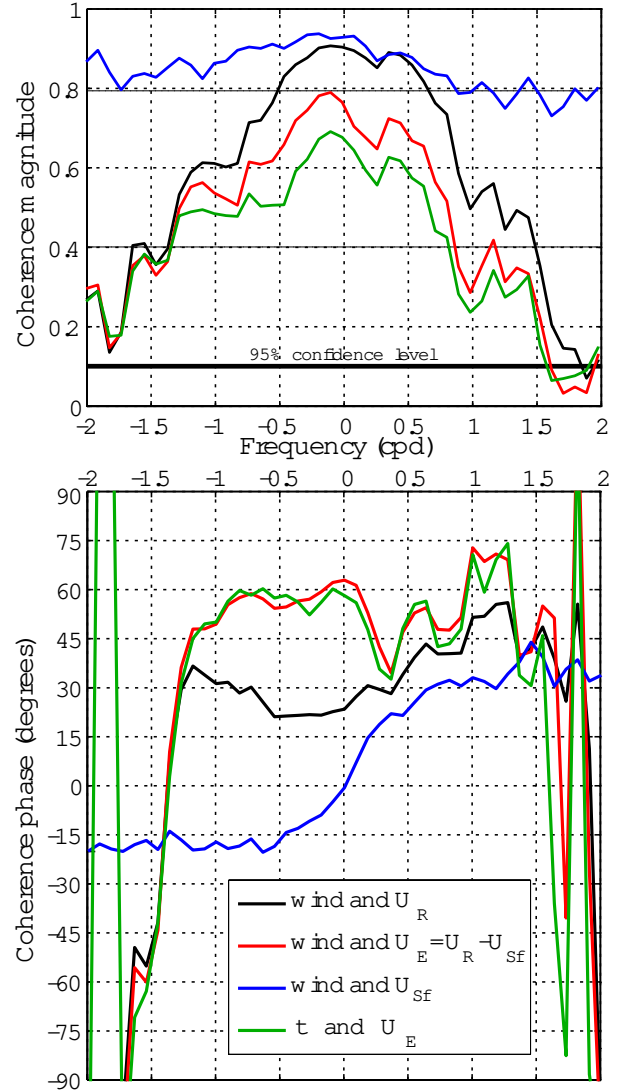}}
%\centerline{\noindent\includegraphics[width=0.7\linewidth]{cospectra_86_revised.pdf}}
\caption{Rotary co-spectra of the wind and wind stress with the
radar-derived current, Stokes drift and Eulerian current. (a)
magnitude and (b) phase. The number of degrees of freedom is 108
at the spectral resolution of 0.09 cpd. Coherence is significant
at the 95\% confidence level for a value of 0.1. Negative and
positive frequencies are clockwise and counter-clockwise polarized
motions, respectively.} \label{cospec}
\end{figure}
%%%%%%%%%%%%%%%%%%%%%%%%%%%%%%%%%%%%%%%%%%%%%%%%%%%%%%%%%%%%%%%%%%%%%%%%%%%%%%%

The radar-measured current vectors $\Ub_R$ have stable directions
relative to the wind, 20 to 40$^\circ$ to the right for $f>-f_I$,
given by their coherence phase (figure 4). The coherence phase of
the Stokes drift increases with frequency. This pattern is typical
of a time lag, that can be estimated to about 1.5~hours, consistent with
the relatively slow response of the wave field compared to the
current. This is rather short compared to the time scale of wave
development, but one should bear in mind that the Stokes drift is
mostly due to short waves that respond faster to the wind forcing
than the dominant waves. Because the wind preferentially turns
clockwise, the Stokes drift is slightly to the left of the wind.
The asymmetry in the phase of $U_{Sf}$ for clockwise and
counter-clockwise motions may be related to varying fetch when
the wind turns.

As expected from the theory by \cite{Gonella1972}, the phase of
the quasi-Eulerian current $U_E$ jumps by about 180$^\circ$ at the
inertial frequency $-f_I$.  In the frequency range from -1.2 to
0.2 cpd, that contains 40\% of the non-tidal signal, $U_E$ is at
an angle between 45 and 60$^\circ$ to the right of the wind. This
conclusion is not much altered when one correlates the Eulerian
current against the wind stress, which, for simplicity is
estimated here with a constant drag coefficient, $\tau = 1.3\times
10^{-3} \mathbf{U}_{10} {U}_{10}$. One may argue that the
theoretical filtering of the Stokes drift is not well validated. A
lower bound on the estimate of $U_{Sf}$ can be given by removing
the contribution from waves shorter than the Bragg waves. This has
very little impact on the estimation of $U_E$.

The observed coherence phases of $U_E$ and $U_{10}$ are similar to
the values given by Gonella (1972, figure 6)\nocite{Gonella1972},
based on the constant eddy-viscosity model of \citet{Ekman1905}, 
but for the current considered at a depth as large as 25\% of the Ekman depth. Since the
radar measurements are representative of the upper 1 meter, and
the Ekman depth is generally of the order of 30~m, it follows that the classical Ekman theory, with a constant eddy viscosity, does not apply here. Instead, this large near-surface deflection  is consistent with model results obtained with a high surface mixing such as
induced by Langmuir circulations \citep{McWilliams&al.1997,Kantha&Clayson2004}, breaking waves
\citep{Craig&Banner1994,Mellor&Blumberg2004,Rascle&al.2006} or both, and consistent with the few observed near-surface velocity profiles \citep{Santala&Terray1992}.

\subsection{Effects of stratification}
Following the theory of Gonella (1972) and the previous
observations by \cite{Price&Sundermeyer1999}, it is expected that
the stratification has a significant effect on the surface
currents. Here we used sea surface temperature time series to
diagnose the presence of a stratification. Because of the strong
vertical mixing year-round at the site of buoy 62069, the
horizontal temperature difference between points A and point 62069
is a good indicator of the vertical stratification at point A.
This temperature difference reaches up to 2$^\circ$C, and was
present in 2006, 2007 and 2008 from early July to late October, as
revealed by satellite SST data. We thus separated the data records
used for the spectral analysis into "stratified" and "homogeneous"
records based on the date of the mid-point in these time series.

These two series show a significant difference (at the 95\%
confidence level) when the spectra are smoothed over 0.3 c.p.d.
bands, with a twice larger response in the cases expected to be
stratified (dashed lines, figure 5) for frequencies in the range
-1.7 to 1.5 c.p.d. Interestingly the transfer functions decrease like $1/(f+\omega)$
from a peak at the inertial frequency $f$, where
$\omega$ is the radian frequency. This decrease is typical of
slab-like behaviors that are expected in mixed layers with a much
larger surface mixing (e.g. Rascle et al. 2006) than typically used with 
Ekman theory, or a mixed layer depth much shallower than the Ekman
depth (Gonella 1972). Ekman theory in unstratified conditions,
that should apply to our winter and spring measurements, would
give a much slower decrease, proportional to $1/\sqrt(f+\omega)$
(Gonella 1972).
%%%%%%%%%%%%%%%%%%%%%%%%%%%%%%%%%%%%%%%%%%%%%%%%%%%%%%%%%%%%%%%%%%%%%%%%%%%%
\begin{figure}[htb!]
\centerline{\noindent\includegraphics[width=\linewidth]{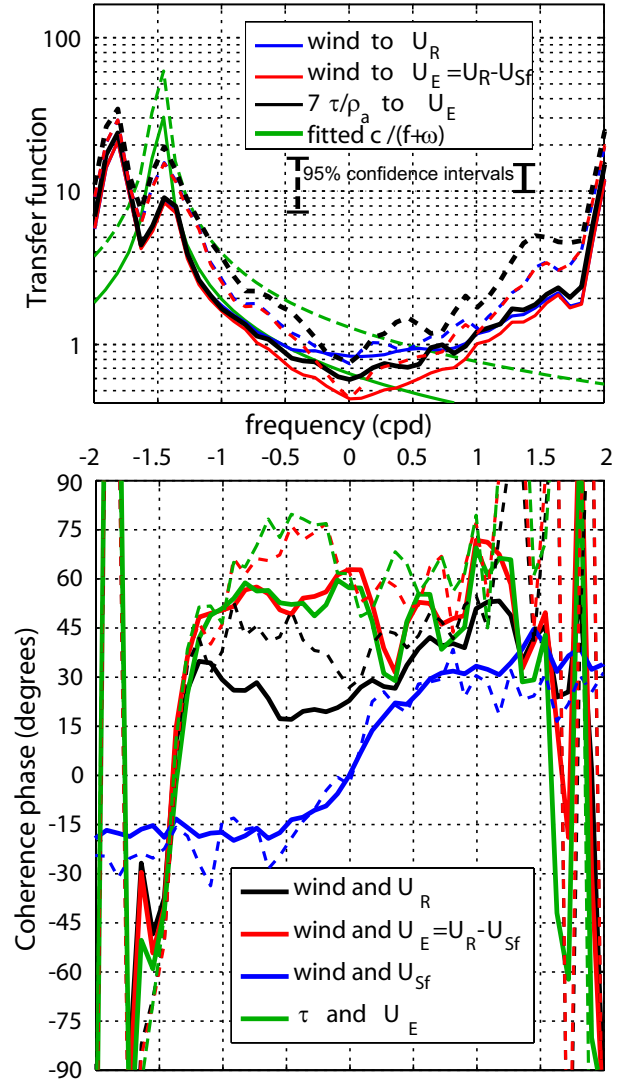}}
%\centerline{\noindent\includegraphics[width=0.6\linewidth]{cospectra_stratif.pdf}}
\caption{Amplitude transfer functions (top) and coherence phases
(bottom) between the wind forcing and the current response. The
dashed lines correspond to records where a stratification is
expected to be important (18 out of 108), and the solid lines
correspond to the other records. Confidence intervals for the two
group of records are shown for the native spectral resolution of
0.09 c.p.d. In order to be at a comparable level the wind stress
was multiplied by 50 before estimating the transfer function. The
two peaks of the transfer functions at +/- 2 cpd are due to the
tidal currents but do not correspond to a causal relationship
between the wind forcing and the current response.}
\label{cospec_strat}
\end{figure}
%%%%%%%%%%%%%%%%%%%%%%%%%%%%%%%%%%%%%%%%%%%%%%%%%%%%%%%%%%%%%%%%%%%%%%%%%%%%%%%
Together with this stronger amplitude of the current response in
stratified conditions, we find a larger deflection angle in the
-0.8 to -0.2 c.p.d. frequency range. This pattern of larger
currents and larger deflection angles in stratified conditions is
consistent with the observations of \cite{Price&Sundermeyer1999},
and the numerical model results by \cite{Rascle&Ardhuin2009}.

\subsection{Relationship between tide-filtered currents and winds}
A proper model for the wind-induced current may be given by the
relationship between the wind speed and wave height, giving the
Stokes drift, and the complex transfer function (transfer function
and phase)  from the wind stress spectrum to the Eulerian current
spectrum, following Gonella (1971) or \cite{Millot&Crepon1981}. Such a model is beyond the scope of 
the present paper.  

Simpler models that would give the current speed and direction as a function of the 
instantaneous wind vector are even less accurate. Because the transfer function is very peaked at the inertial frequency, the current speed may vary widely for a given wind speed. Yet, for practical reasons, there
is a long tradition of directly comparing current and wind
magnitudes and directions for search and rescue operations and ocean
engineering applications. Because of the inertial oscillations,
there is usually a large scatter in the correlation of the current
and wind speed vectors. In order to compare with previous analyses
\citep[e.g.][]{Mao&Heron2008}, we thus perform such a comparison,
after filtering out the dominant tidal current, by taking the
inverse Fourier transform of the current, wind, and Stokes drift
spectra in which the amplitudes of components with frequencies
higher than 1.75~cpd, and the zero frequency, are set to zero.
Again, the Fourier transforms are taken over 264 hours.

We find that the surface Eulerian $U_E$ current lies 40 to
60$^\circ$ to the right of the wind, suggesting that the
near-inertial motions only add scatter to the longer period
motions ($|f|<1.3$ c.p.d.) that were found to have similar
deflection angles. Interestingly, the typical magnitude of $U_E$
decreases from about 0.8\% of $U_{10}$ at low wind to nearly 0.4\%
for high winds. This reduction in the relative magnitude of $U_E$
is accompanied by a reduction of the deflection angle from
65$^\circ$ on average for $U_{10}=3$~m~s$^{-1}$ to 40$^\circ$ for
$U_{10}=15$~m~s$^{-1}$. On the contrary, the Stokes drift
typically increases quadratically with the wind speed. These
observations contradict the usual theoretical statements of \cite{Kirwan&al.1979} and
\cite{Mao&Heron2008}: they concluded that the Stokes drift should be linear and
the Eulerian current should be quadratic in terms of wind speed.
The fact that the Stokes drift is quadratic as a function of the
wind speed is well shown by observed wave spectra in figure C1 and the fitted equation (\ref{Uss_U10}). The
error in \cite{Mao&Heron2008} is likely due to their erroneous assumption
that the Stokes drift is dominated by waves at the peak of the
spectrum. In the analysis of \cite{Kirwan&al.1979} and
\cite{Rascle&al.2006}, the error essentially arises from the
assumed shape of the wave spectrum.

The less-than-linear dependence of $U_E$ on $U_{10}$ contradicts
the usual simple Ekman model for the quasi-Eulerian current, which
would predict a current proportional to the wind stress, and thus
varying as the square or cube of the wind speed. This difference
is likely due to the enhanced mixing caused by breaking waves,
which tends to mix the momentum over a scale of the order of the
windsea wave height, i.e. increasing with the wind speed
\citep{Terray&al.1996,Rascle&al.2006}. Numerical models
without stratification but with a realistic mixing tend to give a
quasi-Eulerian current that increases with wind speed and with the
inverse wave age. Here the stronger winds do not correspond to
very different wave ages, and it is likely that a correlation of
deeper mixed layers with stronger winds is the cause of the
reduction of $U_E$ with increasing wind speed \citep{Rascle&Ardhuin2009}. As a result, the nonlinear 
current response to the wind stress will likely limit the accuracy of models based on transfer functions. 
%%%%%%%%%%%%%%%%%%%%%%%%%%%%%%%%%%%%%%%%%%%%%%%%%%%%%%%%%%%%%%%%%%%%%%%%%%%%
\begin{figure}[htb!]
\centerline{\noindent\includegraphics[width=\linewidth]{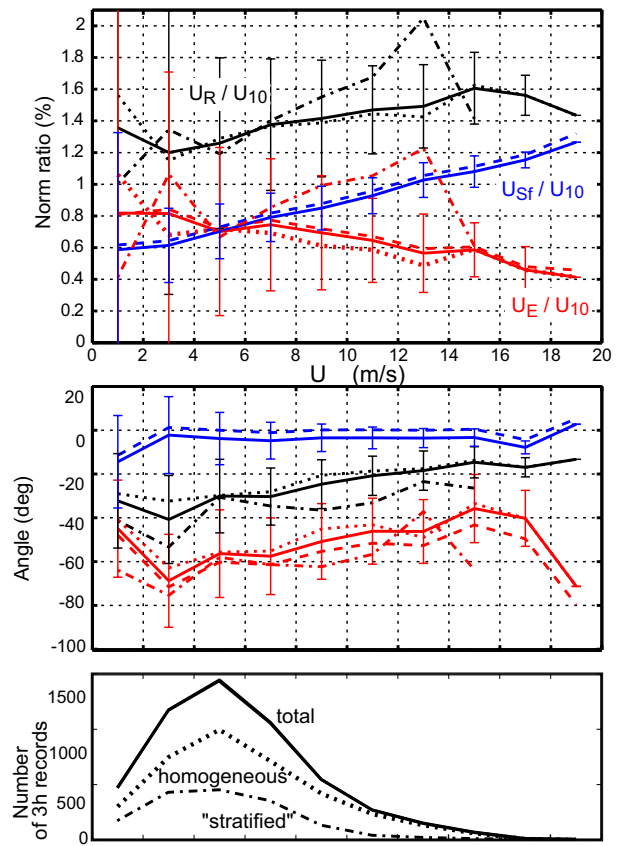}}
%\centerline{\noindent\includegraphics[width=0.6\linewidth]{Ue3_new.pdf}}
\caption{Observed tide-filtered quasi-Eulerian velocity magnitudes, normalized by the wind speed, and directions, 
relative to the wind vector. The linear increase of $U_{Sf}/U_{10}$ with $U_{10}$ is consistent with the quadratic dependence of $U_{Sf}$ on $U_{10}$ given by eq. (\ref{Uss_U10}). The full dataset was binned
according to wind speed. Dash-dotted lines correspond to
stratified conditions only and dotted lines correspond to
homogeneous conditions. The number of data records in each of
these cases is indicated in the bottom panel. The dashed line show
results when $U_{Sf}$ is replaced by $U_{ss}(f_B)$. Error bars
show only $1/2$ of the standard deviation for all conditions
combined, in order to make the plots readable. All time series
(wind, current, $U_{Sf}$ and $U_{ss}$ were filtered in the same
manner for consistency (except for the initial de-tiding applied
only to the current data). The error bars do not represent measurement errors but rather the geophysical 
variability due to inertial motions.} \label{UE3}
\end{figure}
%%%%%%%%%%%%%%%%%%%%%%%%%%%%%%%%%%%%%%%%%%%%%%%%%%%%%%%%%%%%%%%%%%%%%%%%%%%%%%%
%%%%%%%%%%%%%%%%%%%%%%%%%%%%%%%%%%%%%%%%%%%%%%%%%%%%%%%%%%%%%%%%%%%%%%%%%%%%
\begin{figure}[htb!]
\centerline{\noindent\includegraphics[width=\linewidth]{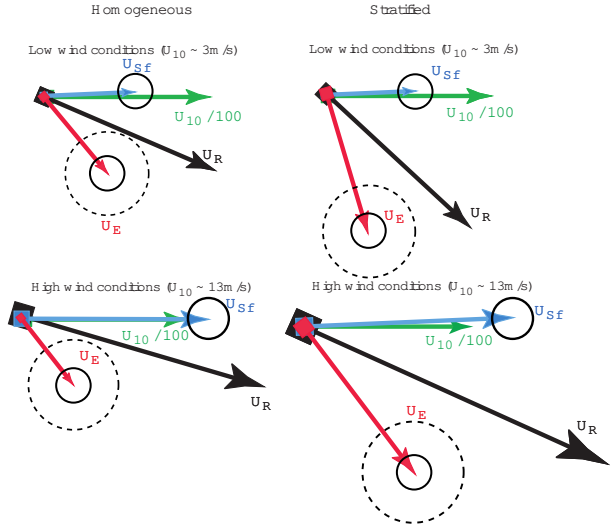}}
%\centerline{\noindent\includegraphics[width=\linewidth]{vectors_revised.pdf}}
\caption{Mean wind-correlated current vectors in low and
high wind conditions, with and without stratification, measured
off the West coast of France with the 12.4 MHz HF radar, based on the results shown in figure 6. $\Ub_R$
is the radar-measured vector, that can be interpreted as a sum of
a quasi-Eulerian current $\Ub_E$, representative of the upper two
meters, and a filtered surface Stokes drift $\Ub_{Sf}$. The full
surface Stokes drift is typically 40\% larger that this filtered
value. Solid circles give the expected error on the mean current components due to biases
in the wave contribution to the radar measurement. The dashed circle show the expected error on the interpretation of $U_E$ as a wind-driven current, based on the ADCP measurements at depth of 60 to 120~m, assuming that the baroclinic part of the geostrophic current is negligible.} \label{UE}
\end{figure}
%%%%%%%%%%%%%%%%%%%%%%%%%%%%%%%%%%%%%%%%%%%%%%%%%%%%%%%%%%%%%%%%%%%%%%%%%%%%%%%

\subsection{Effects of fetch or wave development}
The same analysis was also repeated for other points in the radar
field of view. For example at point B (figure 1), the radar
data quality is generally better, but where the wave model may
have a bias of about 10\% on $U_{ss}$, and the ECMWF wind field
may be less accurate. Point B is relatively sheltered from
Southerly, and North-westerly waves, and the fetch from the East
is 40 km at most. If we assume that the winds are accurate at that
site too, we find that the radar-derived current is weaker
relative to the wind, with $\Ub_R/U_{10}$ typically smaller by
0.2\% point (i.e. a $\sim 15 \%$ reduction) compared to point A.
This appears to be due to a reduction in $U_{Sf}$, which is only
partially compensated for by a small increase in $U_E$. This
difference between $A$ and $B$ nearly vanishes when only Westerly
wind situations are considered (defined by winds 
within 60$^\circ$ from the Westerly direction).

\section{Conclusions}
Using a 2 year time series of HF radar data, and a novel numerical
wave model that is shown to reproduce the observed variability of
the surface Stokes drift with wind speed and wave height, we have
analyzed the wind-driven surface current. When tidal currents are
filtered out, theory predicts that the measured velocities are a
superposition of a filtered Stokes drift $U_{Sf}$  and a
quasi-Eulerian current $U_E$.  With our 12~MHz radar, $U_{Sf}$ is
estimated to be of the order of 0.5 to 1.3\% of the wind speed,
with a percentage that increases linearly with wind speed. These values are a
function of the radar wavelengths and would be larger, by up to
20\%, with higher frequency radars that give currents
representative of a shallower surface layer. The other component
$U_E$ is found to be of the order of 0.6\% of the wind speed, and
lies, in our Northern Hemisphere, at an average 40 to 70 degrees
to the right of the wind, with a large scatter due to inertial
oscillations that may be well modelled using a Laplace transform
of the wind stress \citep{Broche&al.1983}. This large deflection
angle is robustly given by the coherence phase for clockwise
motions in the frequency range from 0 to the inertial frequency.

When instantaneous currents are compared to the wind, the magnitude of $U_E$ appears to decrease with wind speed but it
increases when a stronger stratification is expected (figure 6).
These surface observations correspond to currents in the depth
range 0 to 1.6~m, and confirm previous analysis of deeper
subsurface mooring data. If wind-correlated geostrophic current are negligible in our measurements,  the shape of the classical
picture of the Ekman spiral is not correct, and
the surface layer is much more slab-like than assumed in many
analyses, probably due to the large wave-induced mixing at the
surface \citep{Agrawal&al.1992}.  These findings are summarized in figure 7.

If we neglect the wind-correlated geostrophic currents, which we deem reasonable, and interpret $U_E$ as being purely wind-driven, our observations of $U_E/U_{10}$ at point A are expected to be representative of the open ocean, whereas in coastal areas and small basins, a less developed sea state will lead to a smaller value of $U_{Sf}$ and a larger value of $U_E$, as we observe at point B. Such a generic relationship of $U_E$ and $U_{10}$ is very important
for a proper estimation of the energy flux to the mixed layer.
Besides, on top of the wind stress work on the Ekman current, this energy
flux should be dominated by the dissipation of wave energy induced
by breaking \citep[e.g.][]{Rascle&al.2008}. Also, there is the depth-integrated
Stokes-Coriolis force which is equal to the product of the depth-integrated Stokes transport $\mathbf{M}^w=\rho_w\int \mathbf{U}_s(z) dz$, and the Coriolis parameter $f$. This force is smaller than the depth-integrated Coriolis force by
about a factor of 3 on average \citep{Rascle&al.2008}, but that
may give a comparable work due to the smaller angle between that
force and the quasi-Eulerian current $\widehat{\mathbf{u}}(z)$. The accurate estimation of
the surface Stokes drift using a numerical wave model also opens
the way for a more accurate interpretation of space-borne
measurements of surface currents using Doppler methods, that are
contaminated by a Stokes-like component amplified 10 times or more
\citep{Chapron&al.2005}.

{\it Acknowledgments.} The efforts of Vincent Mariette and Nicolas
Thomas are essential to maintain the radars in proper operating
conditions. Funding for the radar purchase and maintenance was
provided by DGA under the MOUTON project, and funding for the wave
model development was provided under the ECORS project. Florent
Birrien performed the integration of Aaron Roland's routines into
the WAVEWATCH III framework. Wind and wave data were kindly
provided by ECMWF, M{\'e}t{\'e}o-France, and the French Centre
d'Etudes Techniques Maritimes Et Fluviales (CETMEF), and the sea
surface temperature data used to diagnose the presence of a
stratified layer was taken from the ODYSSEA Level 4 global
analysis product, produced as part of the MERSEA Integrated
Project. The SHOM buoy deployments were managed by David Corman
with precious help from Guy Amis.

\appendix
\section{Nonlinear correction for the wave dispersion relation in a random sea state}
Based on the lowest order approximate theory of
\cite{Weber&Barrick1977} for deep water waves with $f\simeq 2 \pi
\sqrt{gk}$, the nonlinear correction to the phase speed of
components with wavenumber $k_B$ and direction $\theta_B$, can be
expressed as an integral over the wave spectrum. Defining
$x=k/k_B$ and $\alpha=\theta-\theta_B$, \cite[][their eq.
A2]{Broche&al.1983} give the following expression,
\begin{eqnarray}
U_{Sf}(k_B,\theta_B) =  \frac{\sqrt{g}}{2} k_B^{3/2}
\int_{0}^\infty \int_0^{2 \pi} F(x,\alpha) E(f,\theta) \mathrm{d}
\theta \mathrm{d} f,\nonumber  \label{UsfA2}
\end{eqnarray}
where, correcting for typographic errors, and using
$y=x^{1/2}=f/f_B$ and $a=\cos \alpha$,
\begin{eqnarray}
F(x,\alpha)&=y\left\{2 a - y+3 x a \right\}
\nonumber \\
& + y \sum_{\varepsilon=\pm 1} \frac{\varepsilon -a}{a_\varepsilon
- \left(1 +\varepsilon y\right)^2}
\nonumber \\
& \times \left\{{\left(y a -x
\right)\left(a_\varepsilon+\left(1+\varepsilon
y\right)^2\right)}/{2}
\right. \nonumber \\
&  \left.+\left(1+ \varepsilon y\right) \left(1+\varepsilon x a +
\varepsilon y\left(x+\varepsilon
a\right)-a_\varepsilon\right)\right\},\nonumber \\
\label{A1corr}
\end{eqnarray}
with
\begin{equation}
a_\varepsilon=\left(1+x^2 + 2 \varepsilon x a \right)^{1/2}.
\end{equation}
These expressions give the correct figures in
\cite{Broche&al.1983}. For $x<1$ one finds that $F(x,0)=4
x^{3/2}$, and for $x>1$, $F(x,0)=4 x^{1/2}$, as previously given
by \cite{Longuet-Higgins&Phillips1962}, \cite{Huang&Tung1976} and
\cite{Barrick&Weber1977}. As commented by \cite{Broche&al.1983},
$F(x,\alpha)\simeq F(x,0) \cos \alpha$, with the largest errors
occurring for $x=1$ where $F(x,\alpha) > F(x,0) \cos \alpha$ for
$| \alpha | < \pi/3$, which, in our case makes $U_{Sf}$ larger by
2 to 5\% than the approximation given by eq. (\ref{Usf}).

\section{Parameterization and numerical settings for the wave
models}
\subsection{Parameterizations}
The implementation of the WAVEWATCH III model used here was ran
with source functions $S_{in}$, $S_{nl}$ and $S_{ds}$
parameterizing the wind input, nonlinear 4-wave interactions and
whitecapping dissipation. An extra additional dissipation term
$S_{db}$ is also included to enhance the dissipation due to wave
breaking in shallow water, based on \cite{Battjes&Janssen1978}.

The parameterization for $S_{nl}$ is taken from
\cite{Hasselmann&al.1985b}, with a minor reduction of the coupling coefficient from $2.78 \times 10^{7}$ to  $2.5 \times 10^{7}$. The parameterizations for $S_{in}$
and $S_{ds}$ are very similar the ones used by
\cite{Ardhuin&al.2008d}, with modifications to further improve the
high frequency part of the spectrum \citep{Filipot&al.2008}.
Namely, the whitecapping dissipation is based on recent
observations of wave breaking statistics \citep{Banner&al.2000},
and swell dissipation \citep{Ardhuin&al.2009b}. These model
settings give the best estimates so far of wave heights, peak and
mean periods, but also of parameters related to the high frequency
tail of the spectrum (appendix C).  The present model results are
thus a significant improvement over the results of
\cite{Bidlot&al.2007} and \cite{Rascle&al.2008}. The physical and
practical motivations for the parameterizations will be fully
described elsewhere, and we only give here a description of their
implementation. We only note for the interested users, that the parameter settings given 
here tend to produce larger negative biasses on $H_s$ for $H_s > 8$~m than the parameterization 
by \cite{Bidlot&al.2007}. Better settings for $H_s$ in extreme waves would be $s_u=0$ and $c_3=0.5$ (see below), but this tends to give too large values of $U_{ss}$, which is why we do not use these settings here.

The parameterization of $S_{in}$ is taken from \cite{Janssen1991}
as modified by \cite{Bidlot&al.2007}, with some further
modifications  for the high frequencies, and the addition of a
wind output term $S_{\mathrm{out}}$ (or "negative wind input")
based on the observations by \cite{Ardhuin&al.2009b}. The source
term is thus
\begin{eqnarray}
S_{\mathrm{in}}\left(f,\theta\right) &=&
\frac{\rho_a}{\rho_w}\frac{\beta_{\mathrm{max}}}{\kappa^2}{\mathrm
e}^{Z}Z^4 \left(\frac{u_\star '}{C}+z_\alpha\right)^2
\nonumber\\
& & \times \cos^2(\theta - \theta_u) \sigma F
\left(f,\theta\right) +
S_{\mathrm{out}}\left(f,\theta\right),\nonumber \\
\label{SinWAM4}
\end{eqnarray}
where $\beta_{\mathrm{max}}$ is a (constant) non-dimensional
growth parameter, $\kappa$ is von K{\'a}rm{\'a}n's constant,
$u_\star$ in the friction velocity in the air, $C$ is the phase
speed of the waves, $\sigma$ is the intrinsic frequency, equal to
$2\pi f$ in the absence of currents, and $ F
\left(f,\theta\right)$ is the frequency-directional spectrum of
the surface elevation variance. In the present implementation the
air/water density ratio is constant. We define $Z=\log(\mu)$ where
$\mu$ is given by Janssen (1991, eq. 16), corrected for
intermediate water depths, so that
\begin{equation}
Z=\log(k z_1)+\kappa/\left[\cos\left(\theta - \theta_u\right)
\left(u_\star ' + z_\alpha \right)\right],\label{Zdef}
\end{equation}
where $z_1$ is a roughness length modified by the wave-supported
stress $\tau_w$, and $z_\alpha$ is a wave age tuning parameter.
The effective roughness $z_1$ is implicitly defined by
\begin{eqnarray}
U_{10}&=&\frac{u_\star}{\kappa} \log\left(\frac{10~\mathrm{m}}{z_1}\right) \\
z_0&=&\max\left\{\alpha_0 \frac{u_\star^2}{g} , 0.0020\right\} \\
z_1&=&\frac {z_0}{  \sqrt{1-\tau_w/\tau}},
\end{eqnarray}
where $\tau$ is the wind stress magnitude, $\tau_w$ is the
wave-supported fraction of the wind stress, $U_{10}$ is the wind
at 10 m height and $g$ is the acceleration of gravity.

 The maximum value
of $z_0$ was added to reduce the unrealistic stresses at high
winds that are otherwise given by the standard parameterization.
This is equivalent to setting a maximum wind drag coefficient of
$2.8\times 10^{-3}$. This, together with the use of an effective
friction velocity $u_\star '(f)$ instead of $u_\star$ in
(\ref{Zdef}) are the only changes to the general form of Janssen's
(1991)\nocite{Janssen1991} wind input. That friction velocity is
defined by
\begin{eqnarray}
\left(u_\star '(f) \right)^2 &=& \left|u_\star^2 \mathrm{e}_{\theta} \right. \nonumber \\
& &\left.- \left|s_u\right| \int_0^f \int_0^{2 \pi}
\frac{S_{in}\left(f',\theta' \right)}{C} \mathrm{e}_{\theta'}
{\mathrm d} f' \mathrm d \theta',\label{ustarp}\right|.\nonumber \\
\end{eqnarray}
Here the empirical factor $s_u=1.0$ adjusts  the sheltering effect
of short waves by long waves adapted from \cite{Chen&Belcher2000},
and helps to reduce the input at high frequency, without which a
balance of source terms would not be possible (except with a very
high dissipation as in Bidlot et al. 2007). This sheltering is
also applied in the precomputed tables that gives the wind stress
as a function of $U_{10}$ and $\tau_w/\tau$
\citep{Bidlot&al.2007}.

The wind output term, is identical to the one used by
\cite{Ardhuin&al.2008d}, based on the satellite observations of \cite{Ardhuin&al.2009b}, with an adjustment to Pacific buoy data. Namely, defining the Reynolds number
Re$=4 u_{\mathrm{orb}} a_{\mathrm{orb}}/\nu_a$, where
$u_{\mathrm{orb}}$ and $a_{\mathrm{orb}}$ are the significant
surface orbital velocity and displacement amplitudes, and $\nu_a$
is the air viscosity, we take, for Re$<10^5$
\begin{equation}
S_{\mathrm{out}}\left(f,\theta\right) = - 1.2
\frac{\rho_a}{\rho_w}\left\{ 2 k \sqrt{2 \nu \sigma}\right\}
F\left(f,\theta\right)\label{Sds_turb} .
\end{equation}
and otherwise
\begin{equation}
S_{\mathrm{out}}\left(f,\theta\right) = -
\frac{\rho_a}{\rho_w}\left\{  16 f_e \sigma^2 u_{\mathrm{orb}} / g
\right\} F\left(f,\theta\right)\label{Sds_visc},
\end{equation}
where
\begin{equation}
f_e = 0.7 f_{e,GM} + \left[0.015 - 0.018 \cos
(\theta-\theta_u)\right]u_\star / u_{\mathrm{orb}},\label{fevar}
\end{equation}
where $f_{e,GM}$ is the friction factor given by Grant and
Madsen's (1979)\nocite{Grant&Madsen1979} theory for rough
oscillatory boundary layers without a mean flow, using a roughness
length adjusted to 0.04 times the roughness for the wind. This
gives a stronger dissipation for swells opposed to winds.

The dissipation term is the sum of the saturation-based term of
\cite{Ardhuin&al.2008d} and a cumulative breaking term
$S_{\mathrm{ds,c}}$ of \cite{Filipot&al.2008}. It thus takes the
form
\begin{eqnarray}
S_{\mathrm{ds}}(f,\theta)& =  \sigma
 C_{\mathrm{ds}} \left\{ 0.25
\left[\max\left\{\frac{B\left(f\right)}{B_r} -
1,0\right\}\right]^2 \right.
\nonumber \\
 &\left. + 0.75 \left[\max\left\{\frac{B'\left(f,\theta \right)}{B_r}
-1
 ,0\right\}\right]^2\right\}\nonumber \\
 &\times  F(f,\theta)  + S_{\mathrm{ds,c}}(f,\theta) \label{Sds_all}.
\end{eqnarray}
where
\begin{equation}
B'\left(f,\theta\right)=
\int_{\theta-80^\circ}^{\theta+80^\circ} k^3 cos^2\left(\theta-
\theta^{\prime}\right) F(f,\theta^{\prime}) C_g / (2 \pi ) \mathrm d
\theta^{\prime} \label{defBofkprime},
\end{equation}
\begin{equation}
B\left(f \right)=\max\left\{B'(f,\theta), \theta \in [0,2
\pi[\right\} \label{defBof},
\end{equation}
and $B_r = 0.0009$ is a threshold for the onset of breaking
consistent with the observations of \cite{Banner&al.2000} and
\cite{Banner&al.2002}, as discussed by
\cite{Babanin&vanderWesthuysen2008}, when including the
normalization by the width of the directional spectrum (here
replaced by the $\cos^2$ factor in eq. \ref{defBofkprime}).

The dissipation constant $C_{\mathrm{ds}}$ was adjusted to
$2.2\times 10^{-4}$ in order to reproduce the directional
fetch-limited data described by \cite{Ardhuin&al.2007}.

The cumulative breaking term represents the smoothing of the
surface by big breakers with celerity $C'$ that wipe out smaller
waves of phase speed $C$ \citep{Babanin&Young2005}. Due to
uncertainties in the estimation of this effect from observations,
we use the theoretical model of \cite{Filipot&al.2008}. Briefly,
the relative velocity of the crests is the norm of the vector
difference, $\Delta_C =\left|\mathbf{C}-\mathbf{C}'\right|$, and
the dissipation rate of short wave is simply the rate of passage
of the large breaker over short waves, i.e. the integral of
$\Delta_C \Lambda(\mathbf{C}) d\mathbf{C}$, where $\Lambda
(\mathbf{C}) d\mathbf{C}$ is the length of breaking crests per
unit surface that have velocity components between $C_x$ and
$C_x+dC_x$, and between $C_y$ and $C_y+dC_y$ \citep{Phillips1985}.
Because there is no consensus on the form of $\Lambda$
\citep{Gemmrich&al.2008}, we prefer to link $\Lambda$ to breaking
probabilities. Based on  Banner et al. (2000, figure 6), and
taking their saturation parameter $\varepsilon$ to be of the order
of $1.6 \sqrt{B}$, the breaking probability of dominant waves
waves is approximately
$P=28.4\left(\max\{\sqrt{B}-\sqrt{B_r},0\}\right)^2$.
In this expression, a division by 2 was included to account for the fact that their breaking probabilities was defined for waves detected using a zero-crossing analysis, which understimates the number of dominant waves because at any given time only one wave is present, and thus low waves of the dominant scale are not counted when shorter but higher waves 
are present. 

Extrapolating this result to higher frequencies, and assuming that
the spectral density of crest length per unit surface
$l(\mathbf{k})$, in the wavenumber spectral space, is
$l(\mathbf{k})= 1/(2\pi^2 k)$, we define a spectral density of
breaking crest length,
$\Lambda(\mathbf{k})=l(\mathbf{k})P(\mathbf{k})$, giving the
source term,
\begin{eqnarray}
S_{\mathrm{ds,c}}(f,\theta) &= -c_3  F \left(f,\theta\right)
\int_0^{0.7f}\int_0^{2\pi}\frac{56.3}{\pi}
\nonumber \\
& \times \max\left\{\sqrt{B(f',\theta'}-\sqrt{B_r},0\right\}
\frac{\Delta_C}{C_g'}\mathrm{d}\theta' \mathrm{d}f'\nonumber\\
\label{Sds_sat_isotropic}.
\end{eqnarray}

The tuning coefficient $c_3$ which was expected to be of order 1,
was here adjusted to 0.4. The resulting model results appear to be
very accurate for sea states with significant wave heights up to
8~m. Larger wave heights are underestimated. Other parameter
adjustments can correct for this defect, e.g. reducing $s_u$ and
increasing $c_3$, but then the Stokes drift may not be so well
reproduced, especially for the average conditions discussed here.
These different possible adjustments and their effects will be
discussed elsewhere.

\subsection{Numerical schemes and model settings}
Spatial advection in the finer model grid is performed using the
explicit CRD-N scheme \citep[Contour integration based Residual
Distribution - Narrow stencil scheme][]{Csik&al.2002} that was
applied to the Wave Action Equation by \cite{Roland2008} and
provided as a module for the WWIII model. The scheme is first
order in time and space, it is conservative and monotone.

All model grids are forced by 6-hourly wind analysis at 0.5 degree
resolution, provided by ECMWF. The model spectral grid has 24
regularly spaced directions, and extends from 0.037 to
$f_{\max}=0.72$~Hz with 32 frequencies exponentially spaced. The
model thus covers the full range of frequencies that contribute
most to the filtered Stokes drift $\Ub_{Sf}$. The usual high
frequency tail proportional to $f^{-5}$ is only imposed for
frequencies larger than the diagnostic frequency $f_d=F
f_{m,0,-1}$, with the mean frequency defined by
$f_{m,0,-1}=\left[\int E(f) / f {\mathrm d}f\right /\int E(f)
{\mathrm d}f]^{-1}$.  Here we take a factor $F=10$, instead of the
usual value of 2.5 \citep{Bidlot&al.2007}, so that $f_d$ is almost
always larger than the model maximum frequency of 0.72~Hz.
Besides, the time step for integration of the source function is
adaptatively refined from 150~s for the local model down to 10~s
if needed, so that virtually no limiter constrains the wave field
evolution \citep{Tolman2002b}.

\section{Model accuracy for relevant parameters} In order to
define the errors on the estimations of $U_{Sf}$ used to determine
the quasi-Eulerian velocity $U_E$ from the radar measurement, it
is necessary to examine the quality of the wind forcing and model
results in the area of interest. The only two parameters that are
measured continuously offshore of the area of interest are the
wave height $H_s$ and mean period $f_{02}$, recorded at buoy
62163, 150 km to the west of point A. $H_s$ and $f_{02}$ can be
combined to give the second moment of the wave spectrum $m_2=(0.25
H_s f_{02})^2$.

%%%%%%%%%%%%%%%%%%%%%%%%%%%%%%%%%%%%%%%%%%%%%%%%%%%%%%%%%%%%%%%%%%%%%%%%%%%%
\begin{figure}[htb!]
\centerline{\noindent\includegraphics[width=0.95\linewidth]{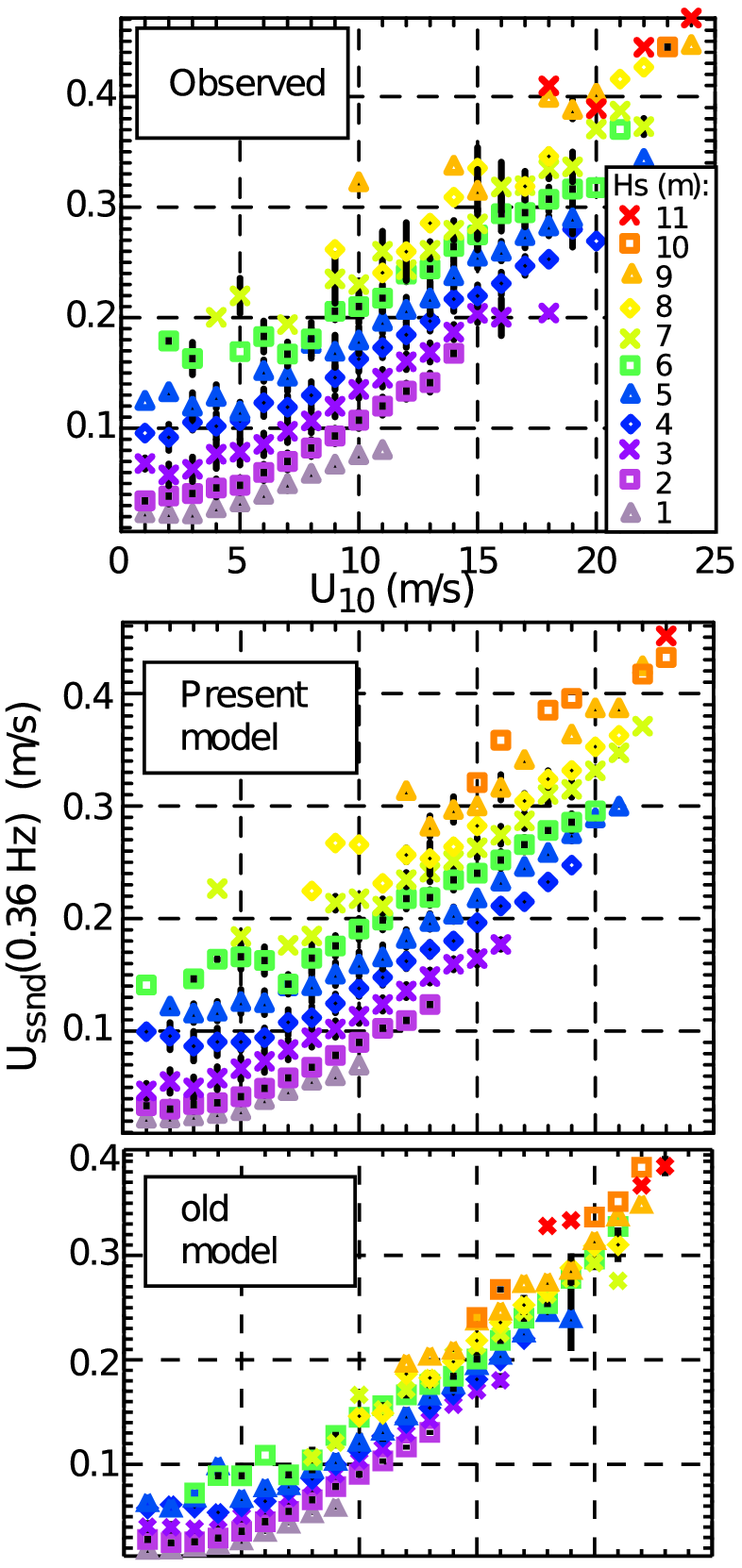}}
%\centerline{\noindent\includegraphics[width=0.5\linewidth]{Uss_U10_HS.pdf}}
\caption{Variation of the wave spectrum third moment, $m_3$
converted to a velocity $U_{\mathrm{ssnd}}=(2 \pi)^3 m_3(f_c)/g$,
that would equal the surface Stokes drift in deep water if all
waves propagated in the same direction. For each data source a
cut-off frequency of $f_c=f_B=0.36$~Hz is taken and the data is
binned wind speed, at 1~m~s$^{-1}$ intervals, and significant wave
height $H_s$ (in colors) at 1~m intervals from 1 to 11~m. The top
panel shows buoy data offshore of Oregon (NDBC buoy 46005), the
middle pannel shows present model results, and the bottom panel
shows results from the same model but using the parameterization
of Bidlot et al. (2007), including a factor $F=2.5$. The vertical
error bars indicate plus and minus half the standard deviation of
the data values in each $(U_{10},H_s)$ class.} \label{fig:Uss_U10}
\end{figure}
%%%%%%%%%%%%%%%%%%%%%%%%%%%%%%%%%%%%%%%%%%%%%%%%%%%%%%%%%%%%%%%%%%%%%%%%%%%%%%%
Because there is no reliable wave measurement with spectral
information in deep water off the French North-East Atlantic
coast, we also use buoy data and model result in a relatively
similar wave environment, at the location of buoy 46005, 650~km
off Aberdeen (WA), on the U.S. Pacific coast. Since this buoy is
not directional we first examine the third moment of the wave
spectrum
\begin{equation}
m_3(f_c) = \int_0^{f_c} f^3 E(f) {\mathrm d}f \label{m3}.
\end{equation}
If waves were all in the same direction, $m_3$ would be
proportional to the Stokes drift $U_{ss}(f_c)$ of waves with
frequency up to $f_c$, as given by eq. (\ref{Uss}). We thus define
a non-directional Stokes drift
\begin{equation}
U_{\mathrm{ssnd}}(f_c)=(2 \pi)^3 m_3(f_c)/g.\label{Ussnd}
\end{equation}
Looking at buoy data we found that
\begin{eqnarray}
U_{\mathrm{ssnd}}(f_c)&\simeq& 5.9\times 10^{-4}
\left[1.25-0.25\left(\frac{0.5}{f_c}\right)^{1.3}\right] U_{10}
\nonumber \\
& \times& \min\left\{U_{10},14.5\right\} + 0.027\left(
H_s-0.4\right),\label{Ussnd_U10} \nonumber\\
\end{eqnarray}
where $f_c$ is in Hertz, $U_{10}$ is in meters per second, and
$H_s$ is in meters.

Taking directionality into account eq. (\ref{Uss}) yields
$U_{\mathrm{ss}}(f_c) \simeq 0.85 U_{\mathrm{ssnd}}(f_c)$, for
typical wave spectra, and the relationship (\ref{Ussnd_U10})
becomes eq. (\ref{Uss_U10}). For buoy 46005, which is a 6~m NOMAD
buoy, and $f_c$ in the range 0.3 to 0.5 Hz, this relationship
gives a root mean square (r. m. s.) error less than
1.0~cm~s$^{-1}$, corresponding to less than 15\% of the r. m. s.
value estimated using eq. (\ref{Ussnd}). This is smaller than the
error of estimates using previous wave models (24\% with the
parameterization by Bidlot et al. 2007), but comparable to the
14.2\% error obtained with the present model. The same analysis was performed, with similar results, for very different sea states recorded by NDBC buoys 51001 (North-East of Hawaii), 41002 (U.S. East coast), 46047 (Tanner Banks, California), 
and 42036 (Gulf of Mexico). 

%%%%%%%%%%%%%%%%%%%%%%%%%%%%%%%%%%%%%%%%%%%%%%%%%%%%%%%%%%%%%%%%%
 \begin{table}[htbp]
  \centering
  \caption[]{Model accuracy for measured wave parameters in various regions of the world ocean. Buoy validation span the entire year 2007,
  except for buoy 62069 for which data covers the time frame 25 January to 20 August 2008, buoy Iroise covers 13 April to 20 May 2004, and JASON 1 data
  corresponds to January to July 2007 for the global validation (JAS-Glo: 393382 data points) and the full year for a box 3$^\circ$ by 4$^\circ$
  centered on 48.5 N and 8 W or 45~N and 128 W.
 (JAS-Gas or JAS-Was: 380 data points).
 Unless otherwise specified by the number in parenthesis, the cut-off frequency is take to be 0.5~Hz, $C$ stands for C-band and $f_B=0.36$~Hz
 corresponds to our 12~MHz HF radar. The normalized bias (NB) is defined as the bias divided by the r.m.s. observed value, while the scatter
 index (SI) is defined as the r.m.s. difference between modeled and observed values, after correction for the bias, normalized by the r.m.s. observed
 value, and $r$ is Pearson's correlation coefficient. Only altimeter data are available at point A but the uniform error pattern and the model consistency suggest that errors at A should be similar to offshore buoy errors such as found at buoy 62163 offshore of A, or at the U.S. West coast buoy 46005. Errors at point B, not discussed here, are expected to be closer to those at the nearshore buoys 62069 and Iroise.}
  \begin{tabular}{lcccc}
\hline
& dataset &  NB(\%)& SI(\%)  & r \\
\hline
2004 & & & & \\
 \hline
$H_s$                    & 62163  &  6.8  & 11.1   & 0.977  \\
$f_{02}$                 & 62163  &  10.4 &  8.8   & 0.907  \\
$H_s$                    & Iroise &  12.8 & 17.4   & 0.975  \\
$f_{02}$                 & Iroise &  -10.0& 11.7   & 0.913  \\
$U_{\mathrm{ssnd}}(f_B)$ & Iroise &  27.2 & 26.9   & 0.968  \\
$U_{ss}(f_B)$            & Iroise &  20.5 & 18.5   & 0.971  \\
\hline
2007/2008& & & & \\
 \hline
$H_{s}$                  & JAS-Glo    &  -0.6 &   11.4 & 0.966 \\
$m_4(C)$                 & JAS-Glo    &   0.6 &   9.1  & 0.939 \\
 \hline
$H_s$                    & 62163  &  -1.4 &   8.8  & 0.985 \\
$f_{02}$                 & 62163  &   6.3 &   7.3  & 0.938 \\
$H_s$                    & 62069  &  10.1 &  14.1  & 0.974 \\
$f_{02}$                 & 62069  &  -7.7 &  11.8  & 0.886 \\
$m_4(f_B)$               & 62069  &  15.8 &  24.1  & 0.955 \\
$U_{\mathrm{ssnd}}(f_B)$ & 62069  &  13.9 &  23.0  & 0.965 \\
$U_{ss}(f_B)$            & 62069  &  11.1 &  21.0  & 0.963 \\
$H_{s}$                  & JAS-Gas  &   -2.6&   8.8  & 0.983 \\
$m4(C)$                  & JAS-Gas  &   1.0 &   6.7  & 0.962 \\
\hline
$H_s$                    & 46005  &   4.9 &  10.2  & 0.975 \\
$f_{02}$                 & 46005  &  -2.8 &   6.6  & 0.931 \\
$m_4(f_B)$               & 46005  &  -5.4 &  13.5 & 0.965 \\
$U_{\mathrm{ssnd}}(f_B)$ & 46005  &  -4.9 &  12.6  & 0.973 \\
$U_{\mathrm{ssnd}}(0.5)$ & 46005  &   6.2 &  12.7  & 0.971 \\
$H_{s}$                  & JAS-Was  &   2.4 &   7.9  & 0.985 \\
$m_4(C)$                 & JAS-Was  &   1.8 &   7.3  & 0.953 \\
 \hline
\end{tabular}
\end{table}
%%%%%%%%%%%%%%%%%%%%%%%%%%%%%%%%%%%%%%%%%%%%%%%%%%%%%%%%%%%%%%%%%%%

 Another source of continous wave measurements is
provided by altimeter-derived $H_s$, which we correct for bias
following \cite{Queffeulou2004}, and fourth spectral moment $m_4$.
The latter is approximately given by \citep{Vandemark&al.2004}
\begin{equation}
m_4 = \frac{0.64 g^2}{(2 \pi)^4 \sigma_0},
\end{equation}
where $\sigma_0$ is the normalized radar cross-section, corrected
for a 1.2 dB bias on the C-band altimeter of JASON in order to fit
airborne observations \citep{Hauser&al.2008}. The model estimation
of $m_4(0.72$~Hz) is extrapolated to C-band by the addition of a
constant $0.011 g^2/(2 \pi)^4$, consistent with the saturation of
the short wave slopes observed by \cite{Vandemark&al.2004}. For
this parameter, the model is found to be very accurate, especially
around the region of interest, relatively more so than on the U.S.
Pacific coast.

These indirect validations suggest that the third spectral moment
including waves up to the Bragg frequency $f_B=0.36$~Hz, which is
proportional to $U_{\mathrm{ssnd}}$, is probably estimated with
bias between -5 and 10\%, and an r.m.s. error less than 20\%. The
bias on the significant wave height appears to increase from
offshore (altimeter and buoy 62163 data), to the coast (buoys
Iroise and 62069), and we attribute this effect to the tidal
currents, not included in the present wave model, and coastal
modifications of the winds that are not well reproduced at this
10-20 km scale by the ECMWF model. Because the chosen area of
interest lies offshore of the area where currents are strongest
(figure 1), we shall assume that, at this site, the model bias on
$U_{ss}(f_B)$ is zero, which appears most likely. Extreme biases of $\pm 10$\% only result in
deflections of $5$ degrees on the diagnosed quasi-Eulerian current
$U_E$.

\bibliographystyle{ametsocjmk}
\bibliography{../references/wave}

\newcommand{\noopsort}[1]{} \newcommand{\printfirst}[2]{#1}
  \newcommand{\singleletter}[1]{#1} \newcommand{\switchargs}[2]{#2#1}
\begin{thebibliography}{81}
\expandafter\ifx\csname natexlab\endcsname\relax\def\natexlab#1{#1}\fi
\expandafter\ifx\csname url\endcsname\relax
  \providecommand{\doi}[1]{doi:\discretionary{}{}{}#1}\else
  \providecommand{\doi}{doi:\discretionary{}{}{}\begingroup
  \urlstyle{rm}\Url}\fi

\bibitem[{Agrawal et~al.(1992)Agrawal, Terray, Donelan, Hwang, Williams,
  Drennan, Kahma, and Kitaigorodskii}]{Agrawal&al.1992}
Agrawal, Y.~C., E.~A. Terray, M.~A. Donelan, P.~A. Hwang, A.~J. Williams,
  W.~Drennan, K.~Kahma, and S.~Kitaigorodskii, 1992: Enhanced dissipation of
  kinetic energy beneath breaking waves. {\it Nature\/}, {\bf 359}, 219--220.

\bibitem[{Ardhuin(2006)}]{Ardhuin2006a}
Ardhuin, F., 2006: Quelles mesures pour la pr{\'e}vision des {\'e}tats de mer
  en zone c{\^o}ti{\`e}re? {\it Communications de l'Atelier Experimentation et
  Instrumentation\/}.
\newline\url{http://www.ifremer.fr/aei2006/resume_long/T1S3/14-aei2006-55.pdf}

\bibitem[{Ardhuin et~al.(2009)Ardhuin, Chapron, and Collard}]{Ardhuin&al.2009b}
Ardhuin, F., B.~Chapron, and F.~Collard, 2009: Observation of swell dissipation
  across oceans. {\it Geophys. Res. Lett.\/}, {\bf 36},
  \doi{\bibinfo{doi}{10.1029/2008GL037030}}, L06607.

\bibitem[{Ardhuin et~al.(2008{\natexlab{a}})Ardhuin, Collard, Chapron,
  Queffeulou, Filipot, and Hamon}]{Ardhuin&al.2008d}
Ardhuin, F., F.~Collard, B.~Chapron, P.~Queffeulou, J.-F. Filipot, and
  M.~Hamon, 2008{\natexlab{a}}: Spectral wave dissipation based on
  observations: a global validation. {\it Proceedings of Chinese-German Joint
  Symposium on Hydraulics and Ocean Engineering, Darmstadt, Germany\/},
  391--400.

\bibitem[{Ardhuin et~al.(2007)Ardhuin, Herbers, Watts, van Vledder, Jensen, and
  Graber}]{Ardhuin&al.2007}
Ardhuin, F., T.~H.~C. Herbers, K.~P. Watts, G.~P. van Vledder, R.~Jensen, and
  H.~Graber, 2007: Swell and slanting fetch effects on wind wave growth. {\it
  J. Phys. Oceanogr.\/}, {\bf 37}, \doi{\bibinfo{doi}{10.1175/JPO3039.1}},
  908--931.

\bibitem[{Ardhuin et~al.(2004)Ardhuin, Martin-Lauzer, Chapron, Craneguy,
  Girard-Ardhuin, and Elfouhaily}]{Ardhuin&al.2004b}
Ardhuin, F., F.-R. Martin-Lauzer, B.~Chapron, P.~Craneguy, F.~Girard-Ardhuin,
  and T.~Elfouhaily, 2004: D{\'e}rive {\`a} la surface de l'oc{\'e}an sous
  l'effet des vagues. {\it Comptes Rendus G\'{e}osciences\/}, {\bf 336},
  \doi{\bibinfo{doi}{10.1016/j.crte.2004.04.007}}, 1121--1130.

\bibitem[{Ardhuin et~al.(2008{\natexlab{b}})Ardhuin, Rascle, and
  Belibassakis}]{Ardhuin&al.2008}
Ardhuin, F., N.~Rascle, and K.~A. Belibassakis, 2008{\natexlab{b}}: Explicit
  wave-averaged primitive equations using a generalized {L}agrangian mean. {\it
  Ocean Modelling\/}, {\bf 20},
  \doi{\bibinfo{doi}{10.1016/j.ocemod.2007.07.001}}, 35--60.

\bibitem[{Babanin and van~der Westhuysen(2008)}]{Babanin&vanderWesthuysen2008}
Babanin, A.~V. and A.~J. van~der Westhuysen, 2008: Physics of saturation-based
  dissipation functions proposed for wave forecast models. {\it J. Phys.
  Oceanogr.\/}, {\bf 38}, 1831--1841.
\newline\url{http://ams.allenpress.com/archive/1520-0485/38/8/pdf/i1520-0485-3%
8-8-1831}

\bibitem[{Babanin and Young(2005)}]{Babanin&Young2005}
Babanin, A.~V. and I.~R. Young, 2005: Two-phase behaviour of the spectral
  dissipation of wind waves. {\it Proceedings of the 5th International
  Symposium Ocean Wave Measurement and Analysis, Madrid, june 2005\/}, ASCE,
  paper number 51.

\bibitem[{Banner et~al.(2000)Banner, Babanin, and Young}]{Banner&al.2000}
Banner, M.~L., A.~V. Babanin, and I.~R. Young, 2000: Breaking probability for
  dominant waves on the sea surface. {\it J. Phys. Oceanogr.\/}, {\bf 30},
  3145--3160.
\newline\url{http://ams.allenpress.com/archive/1520-0485/30/12/pdf/i1520-0485-%
30-12-3145.pdf}

\bibitem[{Banner et~al.(2002)Banner, Gemmrich, and Farmer}]{Banner&al.2002}
Banner, M.~L., J.~R. Gemmrich, and D.~M. Farmer, 2002: Multiscale measurement
  of ocean wave breaking probability. {\it J. Phys. Oceanogr.\/}, {\bf 32},
  3364--3374.
\newline\url{http://ams.allenpress.com/archive/1520-0485/32/12/pdf/i1520-0485-%
32-12-3364.pdf}

\bibitem[{Barrick and Weber(1977)}]{Barrick&Weber1977}
Barrick, D.~E. and B.~L. Weber, 1977: On the nonlinear theory for gravity waves
  on the ocean's surface. {Part II: I}nterpretation and applications. {\it J.
  Phys. Oceanogr.\/}, {\bf 7}, 3--10.
\newline\url{http://ams.allenpress.com/archive/1520-0485/7/1/pdf/i1520-0485-7-%
1-11.pdf}

\bibitem[{Battjes and Janssen(1978)}]{Battjes&Janssen1978}
Battjes, J.~A. and J.~P. F.~M. Janssen, 1978: Energy loss and set-up due to
  breaking of random waves. {\it Proceedings of the 16th international
  conference on coastal engineering\/}, ASCE, 569--587.

\bibitem[{Bidlot et~al.(2007)Bidlot, Janssen, and Abdalla}]{Bidlot&al.2007}
Bidlot, J., P.~Janssen, and S.~Abdalla, 2007: A revised formulation of ocean
  wave dissipation and its model impact. Technical Report Memorandum 509,
  ECMWF, Reading, U. K.

\bibitem[{Broche et~al.(1983)Broche, de~Maistre, and Forget}]{Broche&al.1983}
Broche, P., J.~C. de~Maistre, and P.~Forget, 1983: Mesure par radar
  d\'{e}cam\'{e}trique coh\'{e}rent des courants superficiels engendr\'{e}s par
  le vent. {\it Oceanol. Acta\/}, {\bf 6}, 43--53.

\bibitem[{Chapman et~al.(1997)Chapman, Shay, Graber, Edson, Karachintsev,
  Trump, and Ross}]{Chapman&al.1997}
Chapman, R.~D., L.~K. Shay, H.~Graber, J.~B. Edson, A.~Karachintsev, C.~L.
  Trump, and D.~B. Ross, 1997: On the accuracy of hf radar surface current
  measurements: Intercomparisons with ship-based sensors. {\it J. Geophys.
  Res.\/}, {\bf 102}, 18737--18748.

\bibitem[{Chapron et~al.(2005)Chapron, Collard, and Ardhuin}]{Chapron&al.2005}
Chapron, B., F.~Collard, and F.~Ardhuin, 2005: Direct measurements of ocean
  surface velocity from space: interpretation and validation. {\it J. Geophys.
  Res.\/}, {\bf 110}, doi:10.1029/2004JC002809.

\bibitem[{Chen and Belcher(2000)}]{Chen&Belcher2000}
Chen, G. and S.~E. Belcher, 2000: Effects of long waves on wind-generated
  waves. {\it J. Phys. Oceanogr.\/}, {\bf 30}, 2246--2256.

\bibitem[{Craig and Banner(1994)}]{Craig&Banner1994}
Craig, P.~D. and M.~L. Banner, 1994: Modeling wave-enhanced turbulence in the
  ocean surface layer. {\it J. Phys. Oceanogr.\/}, {\bf 24}, 2546--2559.
\newline\url{http://ams.allenpress.com/archive/1520-0485/24/12/pdf/i1520-0485-%
24-12-2546.pdf}

\bibitem[{Cs{\'i}k et~al.(2002)Cs{\'i}k, Ricchiuto, and
  Deconinck}]{Csik&al.2002}
Cs{\'i}k, {\'A}., M.~Ricchiuto, and H.~Deconinck, 2002: A conservative
  formulation of the multidimensional upwind residual distribution schemes for
  general nonlinear conservation laws. {\it J. Comp. Phys.\/}, {\bf 172},
  286--312.

\bibitem[{Davis(1985)}]{Davis1985}
Davis, R.~E., 1985: Drifter observations of coastal currents during {CODE}: The
  method and descriptive view. {\it J. Geophys. Res.\/}, {\bf 90}, 4741--4755.

\bibitem[{Dobson et~al.(1989)Dobson, Perrie, and Toulany}]{Dobson&al.1989}
Dobson, F., W.~Perrie, and B.~Toulany, 1989: On the deep water fetch laws for
  wind-generated surface gravity waves. {\it Atmosphere Ocean\/}, {\bf 27},
  210--236.

\bibitem[{Ekman(1905)}]{Ekman1905}
Ekman, V.~W., 1905: On the influence of the earth's rotation on ocean currents.
  {\it Ark. Mat. Astron. Fys.\/}, {\bf 2}, 1--53.

\bibitem[{Elipot and Lumpkin(2008)}]{Elipot&Lumpkin2008}
Elipot, S. and R.~Lumpkin, 2008: Spectral description of oceanic near-surface
  variability. {\it Geophys. Res. Lett.\/}, {\bf 35},
  \doi{\bibinfo{doi}{10.1029/2007GL032874}}, L05606.

\bibitem[{Essen(1993)}]{Essen1993}
Essen, H.-H., 1993: Ekman portions of surface currents, as measured by radar in
  different areas. {\it Deut. Hydrogr. Z.\/}, {\bf 45}, 58--85.

\bibitem[{Filipot et~al.(2008)Filipot, Ardhuin, and Babanin}]{Filipot&al.2008}
Filipot, J.-F., F.~Ardhuin, and A.~Babanin, 2008: Param{\'e}trage du
  d{\'e}ferlement des vagues dans les mod{\`e}les spectraux : approches
  semi-empirique et physique. {\it Actes des X{\`e}mes journ{\'e}es G{\'e}nie
  c{\^o}tier-G{\'e}nie civil, Sophia Antipolis\/}, Centre Fran{\c c}ais du
  Littoral.

\bibitem[{Gemmrich et~al.(2008)Gemmrich, Banner, and
  Garrett}]{Gemmrich&al.2008}
Gemmrich, J.~R., M.~L. Banner, and C.~Garrett, 2008: Spectrally resolved energy
  dissipation rate and momentum flux of breaking waves. {\it J. Phys.
  Oceanogr.\/}, {\bf 38}, 1296--1312.
\newline\url{http://ams.allenpress.com/archive/1520-0485/38/6/pdf/i1520-0485-3%
8-6-1296}

\bibitem[{Gonella(1971)}]{Gonella1971}
Gonella, J., 1971: A local study of inertial oscillations in the upper layers
  of the ocean. {\it Deep Sea Res.\/}, {\bf 18}, 776--788.

\bibitem[{Gonella(1972)}]{Gonella1972}
--- 1972: A rotary-component method for analysing meteorological and
  oceanographic vector time series. {\it Deep Sea Res.\/}, {\bf 19}, 833--846.

\bibitem[{Gourrion et~al.(2002)Gourrion, Vandemark, Bailey, and
  Chapron}]{Gourrion&al.2002}
Gourrion, J., D.~Vandemark, S.~Bailey, and B.~Chapron, 2002: Investigation of
  {C}-band altimeter cross section dependence on wind speed and sea state. {\it
  Can. J. Remote Sensing\/}, {\bf 28}, 484--489.

\bibitem[{Grant and Madsen(1979)}]{Grant&Madsen1979}
Grant, W.~D. and O.~S. Madsen, 1979: Combined wave and current interaction with
  a rough bottom. {\it J. Geophys. Res.\/}, {\bf 84}, 1797--1808.

\bibitem[{Gurgel et~al.(1999)Gurgel, Antonischki, Essen, and
  Schlick}]{Gurgel&al.1999}
Gurgel, K.-W., G.~Antonischki, H.-H. Essen, and T.~Schlick, 1999: Wellen radar
  ({WERA}), a new ground-wave based {HF} radar for ocean remote sensing. {\it
  Coastal Eng.\/}, {\bf 37}, 219--234.

\bibitem[{Gurgel and Barbin(2008)}]{Gurgel&Barbin2008}
Gurgel, K.~W. and Y.~. Barbin, 2008: Suppressing radio frequency interference
  in {HF} radars. {\it Sea Technology\/}, {\bf 49}, 39--42.

\bibitem[{Hackett et~al.(2006)Hackett, Breivik, and Wettre}]{Hackett&al.2006}
Hackett, B., {\O}.~Breivik, and C.~Wettre, 2006: Forecasting the drift of
  objects and substances in the ocean. {\it Ocean Weather Forecasting\/}, E.~P.
  Chassignet and J.~Verron, eds., Springer, Netherlands, chapter~23,
  \doi{\bibinfo{doi}{10.1007/1-4020-4028-8}}.

\bibitem[{Hasselmann(1970)}]{Hasselmann1970}
Hasselmann, K., 1970: Wave-driven inertial oscillations. {\it Geophys. Fluid
  Dyn.\/}, {\bf 1}, 463--502.

\bibitem[{Hasselmann et~al.(1985)Hasselmann, Hasselmann, Allender, and
  Barnett}]{Hasselmann&al.1985b}
Hasselmann, S., K.~Hasselmann, J.~Allender, and T.~Barnett, 1985: Computation
  and parameterizations of the nonlinear energy transfer in a gravity-wave
  spectrum. {Part II: P}arameterizations of the nonlinear energy transfer for
  application in wave models. {\it J. Phys. Oceanogr.\/}, {\bf 15}, 1378--1391.

\bibitem[{Hauser et~al.(2008)Hauser, Caudal, Guimbard, and
  Mouche}]{Hauser&al.2008}
Hauser, D., G.~Caudal, S.~Guimbard, and A.~A. Mouche, 2008: A study of the
  slope probability density function of the ocean waves from radar
  observations. {\it J. Geophys. Res.\/}, {\bf 113},
  \doi{\bibinfo{doi}{10.1029/2007JC004264}}, C02006.

\bibitem[{Huang and Tung(1976)}]{Huang&Tung1976}
Huang, N.~E. and C.-C. Tung, 1976: The dispersion relation for a nonlinear
  random gravity wave field. {\it J. Fluid Mech.\/}, {\bf 75}, 337--345.

\bibitem[{Ivonin et~al.(2004)Ivonin, Broche, Devenon, and
  Shrira}]{Ivonin&al.2004}
Ivonin, D.~V., P.~Broche, J.-L. Devenon, and V.~I. Shrira, 2004: Validation of
  {HF} radar probing of the vertical shear of surface currents by acoustic
  {Doppler} current profiler measurements. {\it J. Geophys. Res.\/}, {\bf 101},
  C04003, doi:10.1029/2003JC002025.

\bibitem[{Janssen(1991)}]{Janssen1991}
Janssen, P. A. E.~M., 1991: Quasi-linear theory of wind wave generation applied
  to wave forecasting. {\it J. Phys. Oceanogr.\/}, {\bf 21}, 1631--1642, see
  comments by D. Chalikov, J. Phys. Oceanogr. 1993, vol. 23 pp. 1597--1600.
\newline\url{http://ams.allenpress.com/archive/1520-0485/21/11/pdf/i1520-0485-%
21-11-1631.pdf}

\bibitem[{Jenkins(1987)}]{Jenkins1987}
Jenkins, A.~D., 1987: Wind and wave induced currents in a rotating sea with
  depth-varying eddy viscosity. {\it J. Phys. Oceanogr.\/}, {\bf 17}, 938--951.

\bibitem[{Kahma and Calkoen(1992)}]{Kahma&Calkoen1992}
Kahma, K.~K. and C.~J. Calkoen, 1992: Reconciling discrepancies in the observed
  growth of wind-generated waves. {\it J. Phys. Oceanogr.\/}, {\bf 22},
  1389--1405.
\newline\url{http://ams.allenpress.com/archive/1520-0485/22/12/pdf/i1520-0485-%
22-12-1389.pdf}

\bibitem[{Kantha et~al.(2009)Kantha, Wittmann, Sclavo, and
  Carniel}]{Kantha&al.2009}
Kantha, L., P.~Wittmann, M.~Sclavo, and S.~Carniel, 2009: {\it Geophys. Res.
  Lett.\/}, {\bf 36}, \doi{\bibinfo{doi}{10.1029/2008GL036193}}, L02605.

\bibitem[{Kantha and Clayson(2004)}]{Kantha&Clayson2004}
Kantha, L.~H. and C.~A. Clayson, 2004: On the effect of surface gravity waves
  on mixing in the oceanic mixed layer. {\it Ocean Modelling\/}, {\bf 6},
  101--124.

\bibitem[{Kinsman(1965)}]{Kinsman1965}
Kinsman, B., 1965: {\it Wind waves\/}. Prentice-Hall, Englewood Cliffs, N. J.,
  676 p. Reprinted by Dover Phoenix editions, Mineola, N. Y.

\bibitem[{Kirby and Chen(1989)}]{Kirby&Chen1989}
Kirby, J.~T. and T.-M. Chen, 1989: Surface waves on vertically sheared flows:
  approximate dispersion relations. {\it J. Geophys. Res.\/}, {\bf 94},
  1013--1027.

\bibitem[{Kirwan et~al.(1979)Kirwan, McNally, Pazan, and Wert}]{Kirwan&al.1979}
Kirwan, A.~D., Jr., G.~McNally, S.~Pazan, and R.~Wert, 1979: Analysis of
  surface current response to wind. {\it J. Phys. Oceanogr.\/}, {\bf 9},
  401--412.
\newline\url{http://ams.allenpress.com/archive/1520-0485/9/2/pdf/i1520-0485-9-%
2-401.pdf}

\bibitem[{Le~Boyer et~al.(2009)Le~Boyer, Cambon, Daniault, Herbette, Cann,
  Mari{\'e}, and Morin}]{LeBoyer&al.2009}
Le~Boyer, A., G.~Cambon, N.~Daniault, S.~Herbette, B.~L. Cann, L.~Mari{\'e},
  and P.~Morin, 2009: Observations of the ushant tidal front in september 2007.
  {\it Continental Shelf Research\/}, {\bf 18}, in press.

\bibitem[{Longuet-Higgins and Phillips(1962)}]{Longuet-Higgins&Phillips1962}
Longuet-Higgins, M.~S. and O.~M. Phillips, 1962: Phase velocity effects in
  tertiary wave interactions. {\it J. Fluid Mech.\/}, {\bf 12}, 333--336.

\bibitem[{Madsen(1977)}]{Madsen1977}
Madsen, O.~S., 1977: A realistic model of the wind-induced {E}kman boundary
  layer. {\it J. Phys. Oceanogr.\/}, {\bf 7}, 248--255.

\bibitem[{Mao and Heron(2008)}]{Mao&Heron2008}
Mao, Y. and M.~L. Heron, 2008: The influence of fetch on the response of
  surface currents to wind studied by {HF} ocean surface radar. {\it J. Phys.
  Oceanogr.\/}, {\bf 38}, 1107--1121.
\newline\url{http://ams.allenpress.com/archive/1520-0485/38/5/pdf/i1520-0485-3%
8-5-1107}

\bibitem[{Mariette and Le~Cann(1985)}]{Mariette&LeCann1985}
Mariette, V. and B.~Le~Cann, 1985: Simulation of the formation of the {U}shant
  thermal front. {\it Continental Shelf Research\/}, {\bf 4}, 637.

\bibitem[{McWilliams et~al.(1997)McWilliams, Sullivan, and
  Moeng}]{McWilliams&al.1997}
McWilliams, J.~C., P.~P. Sullivan, and C.-H. Moeng, 1997: Langmuir turbulence
  in the ocean. {\it J. Fluid Mech.\/}, {\bf 334}, 1--30.

\bibitem[{Mellor and Blumberg(2004)}]{Mellor&Blumberg2004}
Mellor, G. and A.~Blumberg, 2004: Wave breaking and ocean surface layer thermal
  response. {\it J. Phys. Oceanogr.\/}, {\bf 34}, 693--698.

\bibitem[{Millot and Cr{\'e}pon(1981)}]{Millot&Crepon1981}
Millot, C. and M.~Cr{\'e}pon, 1981: Inertial oscillations on the continental
  shelf of the {Gulf of Lions}-- observations and theory. {\it J. Phys.
  Oceanogr.\/}, {\bf 11}, 639--657.
\newline\url{http://ams.allenpress.com/archive/1520-0485/11/5/pdf/i1520-0485-1%
1-5-639.pdf}

\bibitem[{Nerheim and Stigebrandt(2006)}]{Nerheim&Stigebrandt2006}
Nerheim, S. and A.~Stigebrandt, 2006: On the influence of buoyancy fluxes on
  wind drift currents. {\it J. Phys. Oceanogr.\/}, {\bf 36}, 1591--1604.

\bibitem[{Pawlowicz et~al.(2002)Pawlowicz, Beardsley, and
  Lentz}]{Pawlowicz&al.2002}
Pawlowicz, R., B.~Beardsley, and S.~Lentz, 2002: Classical tidal harmonic
  analysis including error estimates in {MATLAB using T\_TIDE}. {\it Computers
  and Geosciences\/}, {\bf 28}, 929--937.

\bibitem[{Phillips(1985)}]{Phillips1985}
Phillips, O.~M., 1985: Spectral and statistical properties of the equilibrium
  range in wind-generated gravity waves. {\it J. Fluid Mech.\/}, {\bf 156},
  505--531.

\bibitem[{Pollard(1983)}]{Pollard1983}
Pollard, R.~T., 1983: Observations of the structure of the upper ocean:
  Wind-driven momentum budget. {\it Phil. Trans. Roy. Soc. London A\/}, {\bf
  380}, 407--425.

\bibitem[{Polton et~al.(2005)Polton, Lewis, and Belcher}]{Polton&al.2005}
Polton, J.~A., D.~M. Lewis, and S.~E. Belcher, 2005: The role of wave-induced
  {Coriolis-Stokes} forcing on the wind-driven mixed layer. {\it J. Phys.
  Oceanogr.\/}, {\bf 35}, 444--457.

\bibitem[{Prandle(1987)}]{Prandle1987}
Prandle, D., 1987: The fine-structure of nearshore tidal and residual
  circulations revealed by {H. F.} radar surface current measurements. {\it J.
  Phys. Oceanogr.\/}, {\bf 17}, 231--245.
\newline\url{http://ams.allenpress.com/archive/1520-0485/17/2/pdf/i1520-0485-1%
7-2-231.pdf}

\bibitem[{Price and Sundermeyer(1999)}]{Price&Sundermeyer1999}
Price, J.~F. and M.~A. Sundermeyer, 1999: Stratified {E}kman layers. {\it J.
  Geophys. Res.\/}, {\bf 104}, 20467--20494.

\bibitem[{Queffeulou(2004)}]{Queffeulou2004}
Queffeulou, P., 2004: Long term validation of wave height measurements from
  altimeters. {\it Marine Geodesy\/}, {\bf 27}, 495--510, dOI:
  10.1080/01490410490883478.

\bibitem[{Rascle(2007)}]{Rascle2007}
Rascle, N., 2007: {\it Impact of waves on the ocean circulation (Impact des
  vagues sur la circulation oc{\'e}anique)\/}. Ph.D. thesis, Universit{\'e} de
  Bretagne Occidentale, available at
  http://tel.archives-ouvertes.fr/tel-00182250/.
\newline\url{http://tel.archives-ouvertes.fr/tel-00182250/}

\bibitem[{Rascle and Ardhuin(2009)}]{Rascle&Ardhuin2009}
Rascle, N. and F.~Ardhuin, 2009: Drift and mixing under the ocean surface
  revisited. stratified conditions and model-data comparisons. {\it J. Geophys.
  Res.\/}, {\bf 114}, C02016, doi:10.1029/2007JC004466.

\bibitem[{Rascle et~al.(2008)Rascle, Ardhuin, Queffeulou, and
  Croiz{\'e}-Fillon}]{Rascle&al.2008}
Rascle, N., F.~Ardhuin, P.~Queffeulou, and D.~Croiz{\'e}-Fillon, 2008: A global
  wave parameter database for geophysical applications. part 1:
  wave-current-turbulence interaction parameters for the open ocean based on
  traditional parameterizations. {\it Ocean Modelling\/}, {\bf 25}, 154--171,
  doi:10.1016/j.ocemod.2008.07.006.
\newline\url{http://hal.archives-ouvertes.fr/hal-00201380/}

\bibitem[{Rascle et~al.(2006)Rascle, Ardhuin, and Terray}]{Rascle&al.2006}
Rascle, N., F.~Ardhuin, and E.~A. Terray, 2006: Drift and mixing under the
  ocean surface. a coherent one-dimensional description with application to
  unstratified conditions. {\it J. Geophys. Res.\/}, {\bf 111}, C03016,
  doi:10.1029/2005JC003004.

\bibitem[{Rio and Hernandez(2003)}]{Rio&Hernandez2003}
Rio, M.-H. and F.~Hernandez, 2003: High-frequency response of wind-driven
  currents measured by drifting buoys and altimetry over the world ocean. {\it
  J. Geophys. Res.\/}, {\bf 108}, 3283, doi:10.1029/2002JC001655.

\bibitem[{Roland(2008)}]{Roland2008}
Roland, A., 2008: {\it Development of {WWM II}: Spectral wave modelling on
  unstructured meshes\/}. Ph.D. thesis, Technische Universit{\"a}t Darmstadt,
  Institute of Hydraulic and Water Resources Engineering.

\bibitem[{Santala and Terray(1992)}]{Santala&Terray1992}
Santala, M.~J. and E.~A. Terray, 1992: A technique for making unbiased
  estimates of current shear from a wave-follower. {\it Deep Sea Res.\/}, {\bf
  39}, 607--622.

\bibitem[{Shay et~al.(2007)Shay, Martinez-Pedraja, Cook, and
  Haus}]{Shay&al.2007}
Shay, L.~K., J.~Martinez-Pedraja, T.~M. Cook, and B.~K. Haus, 2007:
  High-frequency radar mapping of surface currents using {WERA}. {\it J. Atmos.
  Ocean Technol.\/}, {\bf 112}, 484--503.

\bibitem[{Smith(2006)}]{Smith2006b}
Smith, J.~A., 2006: Wave-current interactions in finite-depth. {\it J. Phys.
  Oceanogr.\/}, {\bf 36}, 1403--1419.

\bibitem[{Stewart and Joy(1974)}]{Stewart&Joy1974}
Stewart, R.~H. and J.~W. Joy, 1974: {HF} radio measurements of surface
  currents. {\it Deep Sea Res.\/}, {\bf 21}, 1039--1049.

\bibitem[{Terray et~al.(1996)Terray, Donelan, Agrawal, Drennan, Kahma,
  Williams, Hwang, and Kitaigorodskii}]{Terray&al.1996}
Terray, E.~A., M.~A. Donelan, Y.~C. Agrawal, W.~M. Drennan, K.~K. Kahma, A.~J.
  Williams, P.~A. Hwang, and S.~A. Kitaigorodskii, 1996: Estimates of kinetic
  energy dissipation under breaking waves. {\it J. Phys. Oceanogr.\/}, {\bf
  26}, 792--807.

\bibitem[{Tolman(2002)}]{Tolman2002b}
Tolman, H.~L., 2002: Limiters in third-generation wind wave models. {\it Global
  Atmos. Ocean Syst.\/}, {\bf 8}, 67--83.

\bibitem[{Tolman(2007)}]{Tolman2007}
--- 2007: The 2007 release of {WAVEWATCH III}. {\it Proceedings, 10th Int.
  Workshop of Wave Hindcasting and Forecasting, Hawaii\/}.
\newline\url{http://www.waveworkshop.org/10thWaves/Papers/oahu07_Q4.pdf}

\bibitem[{Tolman(2008)}]{Tolman2008}
--- 2008: A mosaic approach to wind wave modeling. {\it Ocean Modelling\/},
  {\bf 25}, \doi{\bibinfo{doi}{10.1016/j.ocemod.2008.06.005}}, 35--47.

\bibitem[{Vandemark et~al.(2004)Vandemark, Chapron, Sun, Crescenti, and
  Graber}]{Vandemark&al.2004}
Vandemark, D., B.~Chapron, J.~Sun, G.~H. Crescenti, and H.~C. Graber, 2004:
  Ocean wave slope observations using radar backscatter and laser altimeters.
  {\it J. Phys. Oceanogr.\/}, {\bf 34}, 2825--2842.

\bibitem[{Wang and Huang(2004)}]{Wang&Huang2004}
Wang, W. and R.~X. Huang, 2004: Wind energy input to the surface waves. {\it J.
  Phys. Oceanogr.\/}, {\bf 34}, 1276--1280.
\newline\url{http://ams.allenpress.com/archive/1520-0485/34/5/pdf/i1520-0485-3%
4-5-1276}

\bibitem[{Weber and Barrick(1977)}]{Weber&Barrick1977}
Weber, B.~L. and D.~E. Barrick, 1977: On the nonlinear theory for gravity waves
  on the ocean's surface. {Part I: D}erivations. {\it J. Phys. Oceanogr.\/},
  {\bf 7}, 3--10.
\newline\url{http://ams.allenpress.com/archive/1520-0485/7/1/pdf/i1520-0485-7-%
1-3.pdf}

\bibitem[{Xu and Bowen(1994)}]{Xu&Bowen1994}
Xu, Z. and A.~J. Bowen, 1994: Wave- and wind-driven flow in water of finite
  depth. {\it J. Phys. Oceanogr.\/}, {\bf 24}, 1850--1866.
\newline\url{http://ams.allenpress.com/archive/1520-0485/24/9/pdf/i1520-0485-2%
4-9-1850.pdf}

\end{thebibliography}

%% ------------------------------------------------------------------------ %%
%
% FIGURES
%
% PLEASE NOTE: WHEN YOU SUBMIT YOUR LATEX FILE TO GEMS, COMMENT OUT ANY COMMANDS
% THAT INCLUDE GRAPHICS (see example below).
%
% ---------------
% ONE-COLUMN figure example
% (For further instructions see FIGURE, PLATE, AND TABLES section at end of file)

%\begin{figure}[htb!]
%\noindent\includegraphics[width=20pc]{samplefigure.eps}}
% \caption{Caption text here}
%\end{figure}
% PLEASE COMMENT OUT YOUR \includegraphics COMMAND AS SHOWN WHEN SUBMITTING
% YOUR ARTICLE.

\end{document}